\documentclass[aps,pre,psfig,twocolumn,showpacs,superscriptaddress,reprint]{revtex4-1}
\usepackage{color}
\usepackage{graphics}
\usepackage{epsfig}
\usepackage[normalem]{ulem} 
\usepackage{cancel}
\usepackage{float}
\usepackage{lineno}
 \usepackage{dcolumn}
 \usepackage{bm}
 \usepackage{amsmath}
 \usepackage{tabularx}
 
\newcommand{\vc}{\mathbf}

\begin{document}
\title{Thermodynamic State Variables in Quasi-Equilibrium Ultracold Neutral Plasma}
\author{Sanat Kumar Tiwari}
\email{sanat-tiwari@uiowa.edu}
\affiliation{Department of Physics and Astronomy, University of Iowa, Iowa City, Iowa 52242, USA}
\author{Nathaniel R.~Shaffer}
\affiliation{Department of Physics and Astronomy, University of Iowa, Iowa City, Iowa 52242, USA}
\author{Scott D.~Baalrud}
\affiliation{Department of Physics and Astronomy, University of Iowa, Iowa City, Iowa 52242, USA}
\date{\today}\emph{•}
\begin{abstract}
The pressure and internal energy of an ultracold plasma in a state of quasi-equilibrium are evaluated using classical molecular dynamics simulations.  
Coulomb collapse is avoided by modeling electron-ion interactions using an attractive Coulomb potential with a repulsive core.  
We present a method to separate the contribution of classical bound states, which form due to recombination, from the contribution of free charges when evaluating these thermodynamic state variables. 
It is found that the contribution from free charges is independent of the choice of repulsive core length-scale when it is sufficiently short-ranged.  
The partial pressure associated with the free charges is found to closely follow that of the one-component plasma model, reaching negative values at strong coupling, while the total system pressure remains positive. 
This pseudo-potential model is also applied to Debye-H\"{u}ckel theory to describe the weakly coupled regime. 
\end{abstract}
\maketitle
\section{Introduction}
\label{intro}
Accurate models for the thermodynamic and transport properties of strongly coupled plasmas are essential for describing their evolution as a continuous fluid~\cite{diaw_15}.
Ultracold neutral plasma (UCP) experiments provide an excellent test bed for validating such models because it is possible to precisely probe them using optical diagnostics in table-top experimental set-ups \cite{strickler_16}.
Verifying models using UCPs  can also advance the understanding of other strongly coupled systems, such as high energy density plasmas \cite{drake2006high,redmer_10,Brueckner_74}, which arise in extreme environments and can be difficult to diagnose precisely.
One of the most intriguing features is that UCPs are electron-ion systems in which each component can be in, or near, the strong coupling regime.
Thus, they can provide insights into two-component physics beyond the reach of the common one-component plasma (OCP) approximation \cite{Baus19801,baus_78}.
In this paper, we develop a method to simulate an electron-ion plasma in a state of quasi-equilibrium using classical molecular dynamics (MD) simulations.
This is applied to evaluate the pressure and internal energy of the system, as well as to distinguish the contributions from free charges and bound states \cite{Bonitz_04}.
These show that the free charge thermodynamics closely resemble predictions from the OCP model, but that classical bound states must also be accounted for to preserve physical limitations such as a positive total pressure. 

UCPs are typically created by the photo-ionization of laser cooled atoms confined in a magneto-optical trap~\cite{killian_prl_99,Killian705,Killian2007}, and can have densities up to $10^{11}\mathrm{cm}^{-3}$.
Ion temperatures at formation range from $\mu$K to mK, and the initial electron temperature typically ranges from 0.1-1 K.
After formation, the plasma components are no longer confined, and the expansion has a cooling effect~\cite{robicheaux_prl_02,killian_prl_01}. 
However, this is overwhelmed by other heating mechanisms.
Both ions and electrons are rapidly heated by disorder induced heating~\cite{guo_pre_10}, and electrons are additionally heated by three-body recombination (3BR) throughout the plasma lifetime~\cite{kuzmin_pop_02}. 
As a result, these are rapidly evolving, partially ionized plasmas with electrons in a weakly to moderately coupled state, and ions in a moderately to strongly coupled state.
 Previous simulation and modeling efforts have largely focused on describing the system evolution, including expansion, disorder induced heating, and eventual recombination to a collapsed neutral-like state~\cite{kuzmin_prl_02,kuzmin_pop_02,mazevet_02}.

Here, we instead focus on developing a method to study the properties of an UCP at fixed conditions, i.e., density and temperature. 
The motivation is to connect theories for thermodynamic and transport properties, which make predictions at fixed conditions, with experiments, which measure these properties over short enough time intervals that the conditions can be considered fixed.
Experiments typically focus on measuring the free charges~\cite{strickler_16}. 
It is interesting from a theoretical viewpoint -- and necessary for comparison with experiment -- that one separates the bound state contributions from the free charge contributions when describing transport or thermodynamic properties. 
The primary challenge is that the equilibrium state of the system is a recombined neutral gas \cite{kuzmin_pop_02,fortov_2000,klimontovich1967}. 
A successful model must somehow limit the recombination so that a free charge population remains, but do so in a way that one can connect that simulated equilibrium state with an interval of time during the evolution of the plasma in an experiment. 

To accomplish this, we model electron-electron and ion-ion interactions with the Coulomb potential and electron-ion interactions with a pseudo Coulomb potential that also includes a repulsive core
\begin{subequations}
\label{eq:pots}
 \begin{eqnarray}
 && v_{ee} = v_{ii} = \frac{e^2}{r}  \\
 && v_{ei} = -\frac{e^2}{r} \left[ 1 - \exp{\left(-\frac{r^2}{(\alpha a)^2}\right)} \right] .
 \label{pot_md}
 \end{eqnarray}
 \end{subequations}
Here, $r$ is the separation between two charged particles, $e$ is the electron charge, $a=(3/4\pi n)^{1/3}$ is the average interparticle spacing based on the total number density $n=n_e + n_i$, and $\alpha$ is an adjustable parameter that sets the e-i repulsion length scale.
Simulations were conducted in a periodic box with both electrons and ions held to the same fixed temperature using a Nos\'{e}-Hoover thermostat 
\cite{evans_85}. 
Due to computing constraints, the ion mass was set to be 10 times the electron mass.
Since mass does not influence the equilibrium properties of the system, which are the focus of this work, this reduced mass is inconsequential. 
Electrons are hotter than ions in real UCP experiments, but we concentrate on equilibrium here because our interpretation of data will utilize aspects of equilibrium statistical mechanics.  
Future work will extend the model to treat unequal electron and ion temperatures. 

The electron-ion potential in Eq.~\eqref{pot_md} is similar to the Kelbg potential used to model dense, degenerate plasmas~\cite{deutsch_77,kelbg_63}.
However, an important difference arises here. 
In dense plasmas, the length scale $\alpha a$ is associated with the de Broglie wavelength characterizing quantum mechanical diffraction. 
At dense plasma conditions, the de Broglie wavelength is of the same order as the inter-particle spacing, so $\alpha a$ is of order unity. 
As a result, $\alpha$ significantly influences the predicted thermodynamic properties and transport rates. 
In contrast, in a UCP the de Broglie wavelength is orders of magnitude smaller than $a$. 
In our model, $\alpha$ is a model parameter that does not represent a physical scale. 

The main idea behind this model is that as $\alpha$ decreases, the properties of the free charge components of the system asymptote to values that are independent of $\alpha$. 
Hence, these asymptotic values represent the state of the charged components at fixed conditions. 
The main result of this paper is the demonstration of this asymptotic plateau in the  pressure and internal energy as the parameter $\alpha$ is reduced.  
What does change as $\alpha$ shrinks is the fraction of the plasma in a bound state. 
Decreasing $\alpha$ increases the depth of the potential well in the electron-ion interaction, resulting in more classically bound pairs, or clusters. 
We observe that the bound state population has a lower temperature than the free population.
This, along with a decreasing fraction of free charged states, leads to a slight slope in the thermodynamic variable profiles as $\alpha$ decreases.
Nevertheless, the model provides a means to access properties of the charged particles (plasma) at fixed conditions via the asymptotic values obtained at small $\alpha$, while also providing a means of controlling the bound state fraction.  

Interpretation of the data requires a means to separate bound states from free charges. 
Here, we calculate the electron-electron, ion-ion and electron-ion radial distribution functions, $g_{ij}(r)$, and apply a simple model based on an energy argument to separate free and bound states. 
The pressure and internal energy are then computed directly from the radial distribution functions. 
The results provide a proof of principle of this technique. 
Future developments may address methods to directly separate free and bound states in the simulations, as well as to treat non-equilibrium systems that more closely represent experimental conditions. 
 
This paper is organized as follows. 
Section \ref{th_weak} applies the model to Deybe-H\"{u}ckel theory, which treats weakly coupled plasmas. 
This serves to demonstrate key aspects of the model using a familiar analytic formalism. 
Section \ref{sim_mod} provides details of the MD simulations. 
Section \ref{th_strong} presents the results of applying the model using MD simulations at strongly coupled conditions.  
Finally, we conclude and provide some future prospects in Sec.~\ref{concl}. 

\section{Weakly coupled plasma}
\label{th_weak}

At equilibrium, the coupling strength can be quantified by the Coulomb coupling parameter 
\begin{equation}
\Gamma = \frac{e^2/a}{k_BT} ,
\end{equation}
which is the ratio of the Coulomb potential energy at the average interparticle spacing to the average kinetic energy. 
Properties of weakly coupled plasmas, $\Gamma \ll 1$, are well described by models based on a series of binary interactions between particles. 
In this section, we first revisit aspects of two- and three-body interactions that will be useful for interpreting the more complex N-body simulations in Sec.~\ref{th_strong}. 
We also apply the model potentials to Debye-H\"{u}ckel theory, demonstrating their essential features: the separation of bound and free states and the asymptotic values of the free-charge thermodynamic state variables as $\alpha$ is reduced. 

\subsection{Classical bound states}
\label{cl_bd}
\begin{figure}
 \includegraphics[height = 4.0cm,width = 9.0cm]{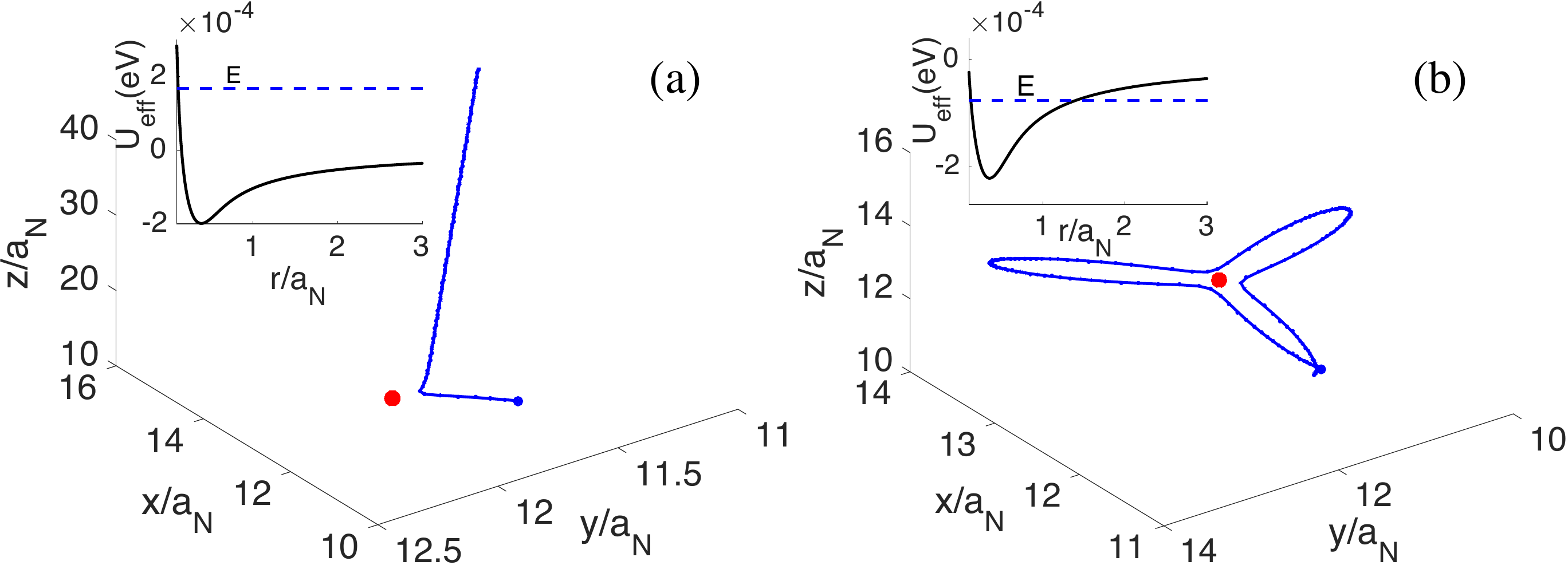}
 \caption{Trajectories of an electron (blue) and ion (red) interacting through the Coulomb potential at conditions representing (a) free scattering and (b) a bound state.  Arrows show the direction of electron motion from its starting point (blue dot). 
  The ion to electron mass ratio is chosen to be $10^5$ for this case. Distances are in units of $a_N = 10^{-5}$m. 
  }
 \label{fig_bd}
\end{figure}
%

Binary encounters between electrons and ions can be classified as either free or bound.
Since the effective potential, $U_{\textrm{eff}}(r) = v_{ei}(r) + l^2/(2m_{ei}r^2)$, has a global minimum, the sign of total energy, $E = m_{ei} u^2/2 + v_{ei}(r)$, of the e-i pair determines whether the orbit is bound or free~\cite{taylor_04}.
Here, $m_{ei} = m_e m_i/(m_e + m_i)$ is the reduced mass, $u = | \vc{v}_e - \vc{v}_i|$ is the relative initial particle speed, and $l$ is the angular momentum.
Figure~\ref{fig_bd} shows an example of each type of interaction for an electron-ion pair interacting via the Coulomb potential. 
In Figs.~\ref{fig_bd}a and b, the initial conditions are such that $E > 0$ and $E < 0$ respectively, resulting in free and bound orbits.

\subsection{Three-body interactions}
\label{3br_cl}
\begin{figure}
 \includegraphics[width=8cm]{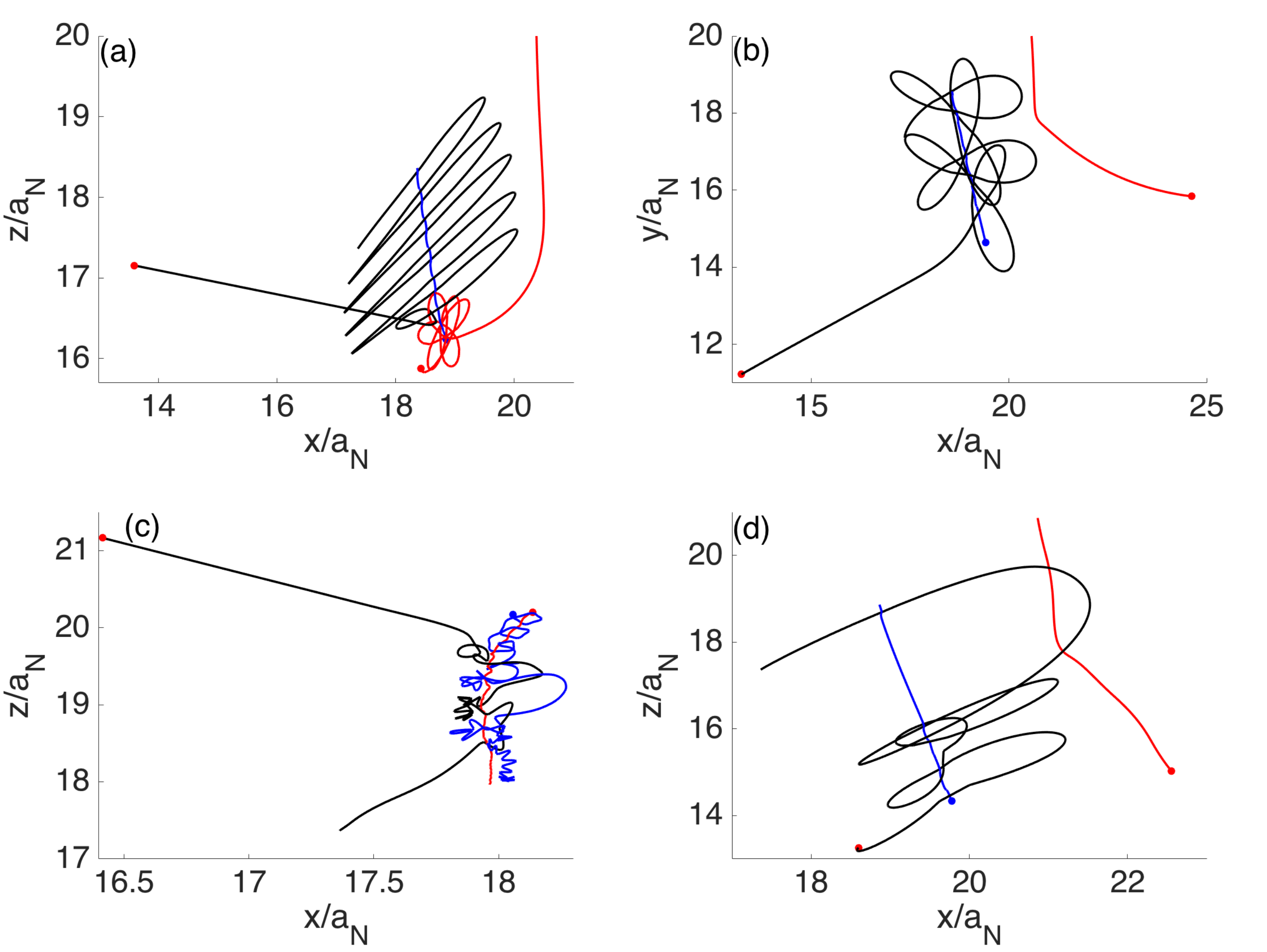}
 \caption{Trajectories demonstrating various outcomes of three-body interactions between two electrons (red and black) and one ion (blue): (a) A classical electron-ion bound state interacts with an energetic electron, which frees the previously bound electron and forms a loosely bound state with the ion. (b) A classical electron ion-bound state interacts with an energetic electron, resulting in all free states. (c) An external electron interacts with a loosely bound state, gaining kinetic energy from the interaction and causing the bound pair to become more tightly bound. (d) An ion and two electrons all begin in a free state but form a bound pair via the three-body recombination. The ion to electron mass ratio is chosen to be 100 for this case. Distances are normalized with $a_N = 1.3366 \times 10^{-5}$m.
 }
 \label{fig_3BR}
\end{figure}

Binary collisions alone do not allow for the formation of bound states from free states, since the total energy of the binary pair is fixed.
However, this can change if a third particle is present.
Four types of three-body interactions are pictured in Figure~\ref{fig_3BR}.
Figure~\ref{fig_3BR}d illustrates the interaction of two electrons and an ion -- all initially free -- to form a bound electron-ion pair.
The reduced potential energy of the newly bound pair is transferred to the second electron as additional kinetic energy.
This is a classical realization of three-body recombination (3BR), which is an important heating mechanism in ultracold plasmas~\cite{kuzmin_prl_02,fletcher_prl_07,killian_prl_01}.
At thermal equilibrium, the formation of bound states is balanced by the reciprocal process, classical impact ionization, which is pictured in Figure~\ref{fig_3BR}b.
The net result of these three-body interactions is that the bound pairs are less energetic than free particles, leading to overall heating of the free charges in the plasma, especially the electrons. 
At equilibrium, the bound subset may have a lower temperature than the free population. 
This will be discussed further in Sec.~\ref{tbr_sec}. 

\begin{figure*}
 \includegraphics[height = 5.0cm,width = 16.0cm]{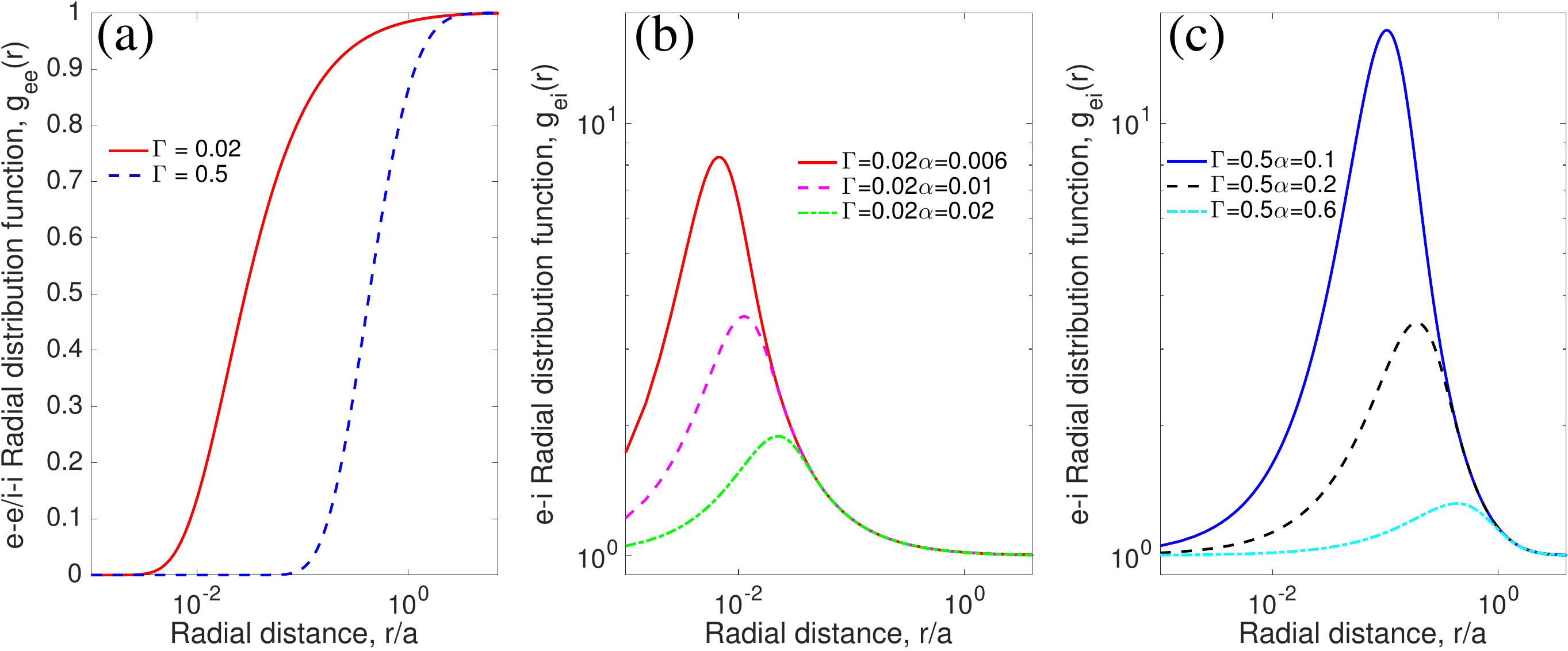}
 \caption{Radial distribution functions at weakly coupled conditions obtained from Eq.~\eqref{mean_pot}. (a) Electron-electron/ion-ion RDFs for $\Gamma=0.02, 0.5$. 
 (b) Electron-ion RDFs for $\Gamma = 0.02$ and various $\alpha$. (c) Electron-ion RDFs for $\Gamma = 0.5$ and various $\alpha$. Like species RDFs ($g_{ee/ii}$) have no $\alpha$ dependence as they interact through the bare Coulomb potential.}
 \label{fig_rdfsw}
\end{figure*}

\subsection{Radial distribution functions}
\label{rdf_dh}
The radial distribution function represents the density profile surrounding individual charged particles. 
It is also related to the potential of mean force, which is the potential obtained when taking two particles at fixed positions and averaging over the positions of all other particles \cite{hill_2012}
\begin{eqnarray}
\vc{F}_{12} &=& \int \biggl[ - \nabla_{\vc{r}_1} U(\vc{r}_1,  \ldots , \vc{r}_N) \biggr] \frac{e^{-U/k_BT}}{\mathcal{Z}} d\vc{r}_3 \ldots d\vc{r}_N  \label{eq:f12} \\ \nonumber
&=& - k_BT \nabla_{\vc{r}_1} \ln g(|\vc{r}_1 - \vc{r}_2|) \equiv - \nabla_{\vc{r}_1} \phi (\vc{r}_1 - \vc{r}_2) .
\end{eqnarray}
Here, $g(r)$ is the radial distribution function, $\phi$ is the potential of mean force, $\mathcal{Z} = \int \exp(-U/k_BT) d\vc{r}_1\ldots{}d\vc{r}_N$ is the configurational integral and $U \equiv \sum_{i,j} v(|\vc{r}_i - \vc{r}_j|)$.  

In weakly coupled plasmas, the potential of mean force is the Debye-H\"uckel potential with a screening length equal to the total Debye length.
 This can be obtained using a standard fluid approach with a Boltzmann distribution of electrons and ions \cite{chen1974}, or from the potential of mean force computed from the weakly coupled limit of the hypernetted-chain (HNC) approximation ($\phi/k_BT \ll 1$)~\cite{baalrud_pop_14}. 
For the bare potentials in Eq.~\eqref{eq:pots}, the associated weakly coupled limit of the potentials of mean force are
\begin{subequations}
\label{mean_pot}
\begin{eqnarray}
\frac{\phi_{ii}(r)}{k_BT} &=& \frac{\phi_{ee}(r)}{k_BT} =  \frac{\Gamma}{r/a} \exp{\left(-\sqrt{3\Gamma} r/a  \right)} \\
\frac{\phi_{ei}(r)}{k_BT} &\simeq&  - \frac{\phi_{ii}(r)}{k_BT} \left\lbrace1 - \exp [-\left(r/\alpha a \right)^2 ] \right\rbrace .
\end{eqnarray}
\end{subequations}
The expression for $\phi_{ei}(r)$ relies on a scale separation between the repulsive core and screening length ($\alpha a \ll \lambda_D$). 
The RDFs can be obtained directly from Eq.~\eqref{mean_pot} via their association with the potential of mean force $g_{ij}(r) = \exp(-\phi_{ij}/k_BT)$.
Note that since both species are assumed to have the same temperature, $g_{ee} = g_{ii}$ and $g_{ei} = g_{ie}$. 
Figure~\ref{fig_rdfsw} shows the RDFs for (a) electron-electron (or ion-ion) pairs with coupling strength $\Gamma = 0.02$ and $\Gamma=0.5$ (red and blue lines respectively), (b) electron-ion pairs with $\Gamma = 0.02$ and (c) electron-ion pairs with $\Gamma = 0.5$.
The electron-ion RDFs ($g_{ei}$) clearly show a peak at the location $\alpha a$ with the amplitude of this peak increasing sharply either as $\alpha$ decreases or as the coupling strength increases.  
These peaks represent the classical bound states that form in the  potential well at separation $\alpha a$. 

\begin{figure}
 \includegraphics[width = 7.0cm]{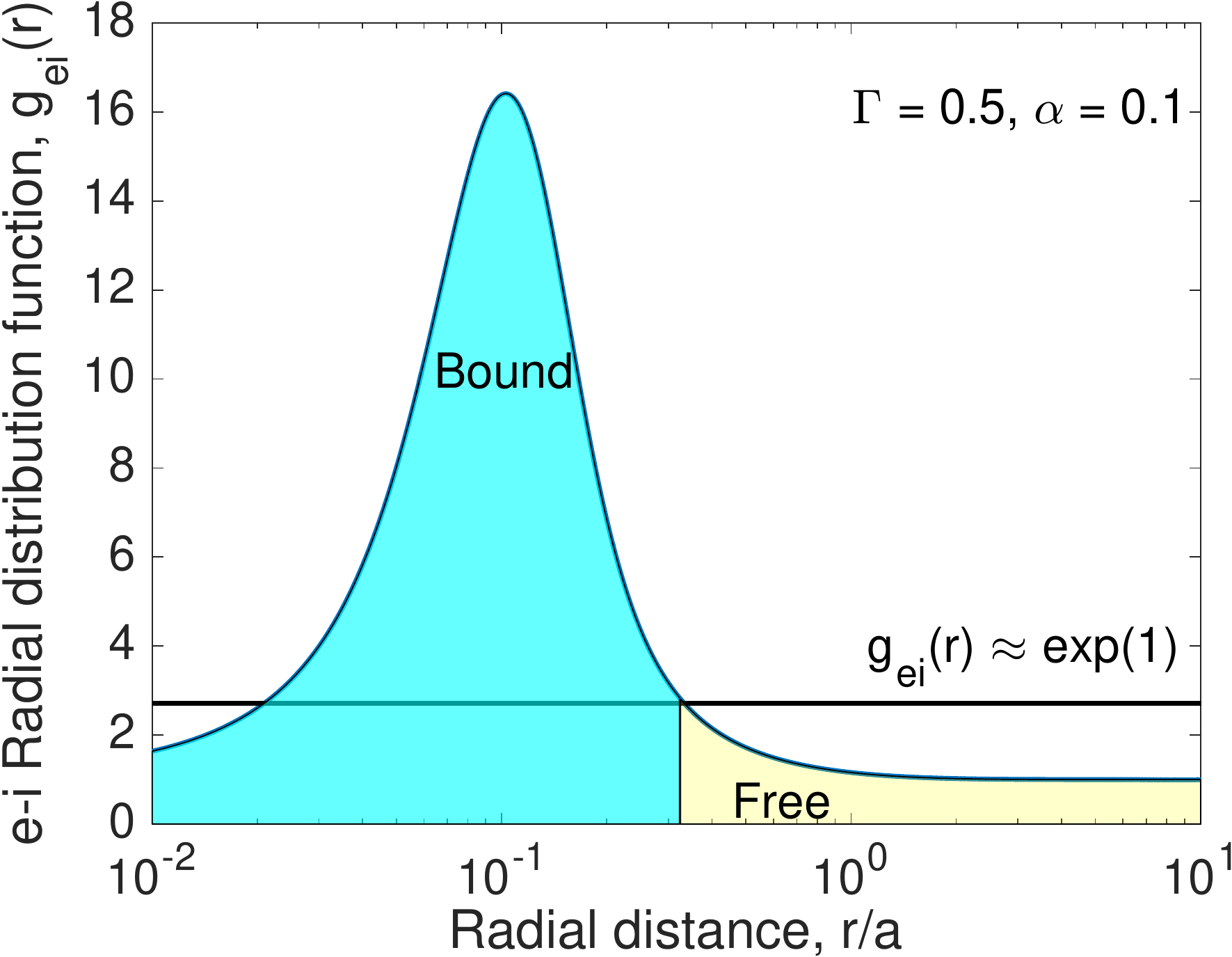}
 \caption{Free (yellow) and bound state (cyan) contributions to $g_{ei}(r)$ with $\Gamma=0.5$ and $\alpha=0.1$. The horizontal and vertical lines delineate $g_{ei}=1$ and $r_c$, respectively.}
 \label{bd_free}
\end{figure}

Next, we discuss a method to distinguish contributions due to free and bound charges in the RDFs, which will later be used to distinguish the contributions of each population to the thermodynamic state variables. 
As discussed in Sections~\ref{cl_bd} and~\ref{3br_cl}, the condition for an e-i pair to be bound is $E < 0$, which can occur as the result of interaction with a third particle. 
In a many-body picture, the potential of mean force models the effective interaction energy of an e-i pair in the presence of the surrounding plasma. 
Applying this to the condition for bound states from Sec.~\ref{3br_cl} suggests that particle interactions for which $|\phi_{ei}(\vc{r}_{12})| > k_BT$ are expected to be bound and those with $|\phi_{ei}(\vc{r}_{12})| < k_BT$ free. 
We use this as a criterion to separate $g_{ei}(r)$ into free and bound contributions according to
\begin{equation}
\max{\lbrace g_{ei}^{\textrm{free}} \rbrace} = \exp{\left(1 \right)}~. \label{eq:g_max}
\end{equation}
In other words, a critical distance $r_c$ defined by $|\phi_{ei}(r_c)|=k_BT$ delineates the separation between bound and free populations: Particles in the region $r>r_c$ are free and those with $r<r_c$ bound. Figure~\ref{bd_free} provides an example for $\Gamma = 0.5$ and $\alpha = 0.1$, showing the separation between bound and free contributions to the radial density profile.

\begin{figure}
 \includegraphics[width = 8.0cm]{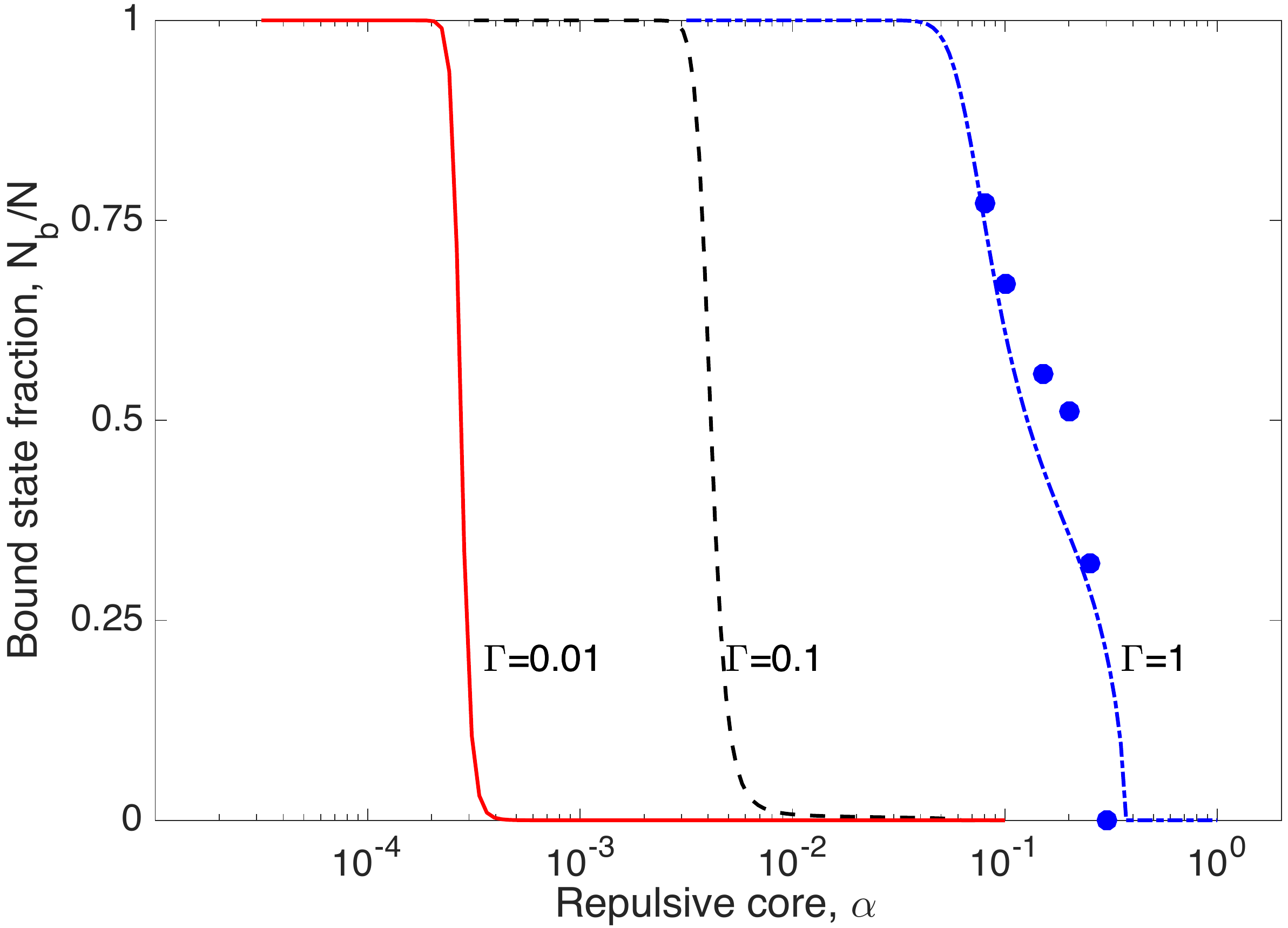}
 \caption{Fraction of bound states with respect to repulsive core parameter $\alpha$ obtained from Eqs.~(\ref{eq:g_max}) and (\ref{eq:bound_frac}). The lines use $g_{ie}$ from Debye-H\"uckel theory, Eq.~(\ref{mean_pot}), and filled blue circles use $g_{ie}$ from MD simulations for $\Gamma = 1$.}
 \label{bd_frac_dh}
\end{figure}

The bound state fraction can be estimated directly from the e-i RDFs by taking the ratio of the number of bound particles to the total number of particles
\begin{equation}
  \label{eq:bound_frac}
  \frac{N_b}{N} = \frac{\int_0^{r_c} \left[g_{ei}(r)-1\right]d\vc{r}}{\int_0^{\infty} \left[g_{ei}(r)-1\right] d\vc{r}}.
\end{equation}
Figure~\ref{bd_frac_dh} illustrates how the fraction of bound states varies with the repulsive core parameter $\alpha$.
At a given coupling strength, there is a transition regime where the bound state fraction increases sharply. The upper edge of this region indicates a nearly recombined plasma (i.e., classical neutral gas) while the lower edge indicates a fully ionized plasma. The transition is observed to occur when 
$\alpha \simeq 0.05 \Gamma$ based on this data in the range $\Gamma = 0.01 - 1$.

\subsection{Excess pressure}
\label{ex_pr}
At equilibrium, the pressure can be computed directly from the RDFs. 
It consists of an ideal component and an excess component: $P= P_{\mathrm{ideal}} + P_{\mathrm{ex}}$, where $P_{\mathrm{ideal}} = n k_BT$, and the excess pressure is \cite{hill_2012,hansen_pra_81}
\begin{equation}
P_{\mathrm{ex}} = -\frac{2}{3} \pi \sum_{i,j} n_i n_j \int_0^\infty v^\prime_{ij}(r) g_{ij}(r) r^3 dr~,
\label{pex_we}
\end{equation}
where $v_{ij}^\prime$ denotes the radial derivative of the bare potentials.

\begin{figure}
 \includegraphics[width = 7.0cm]{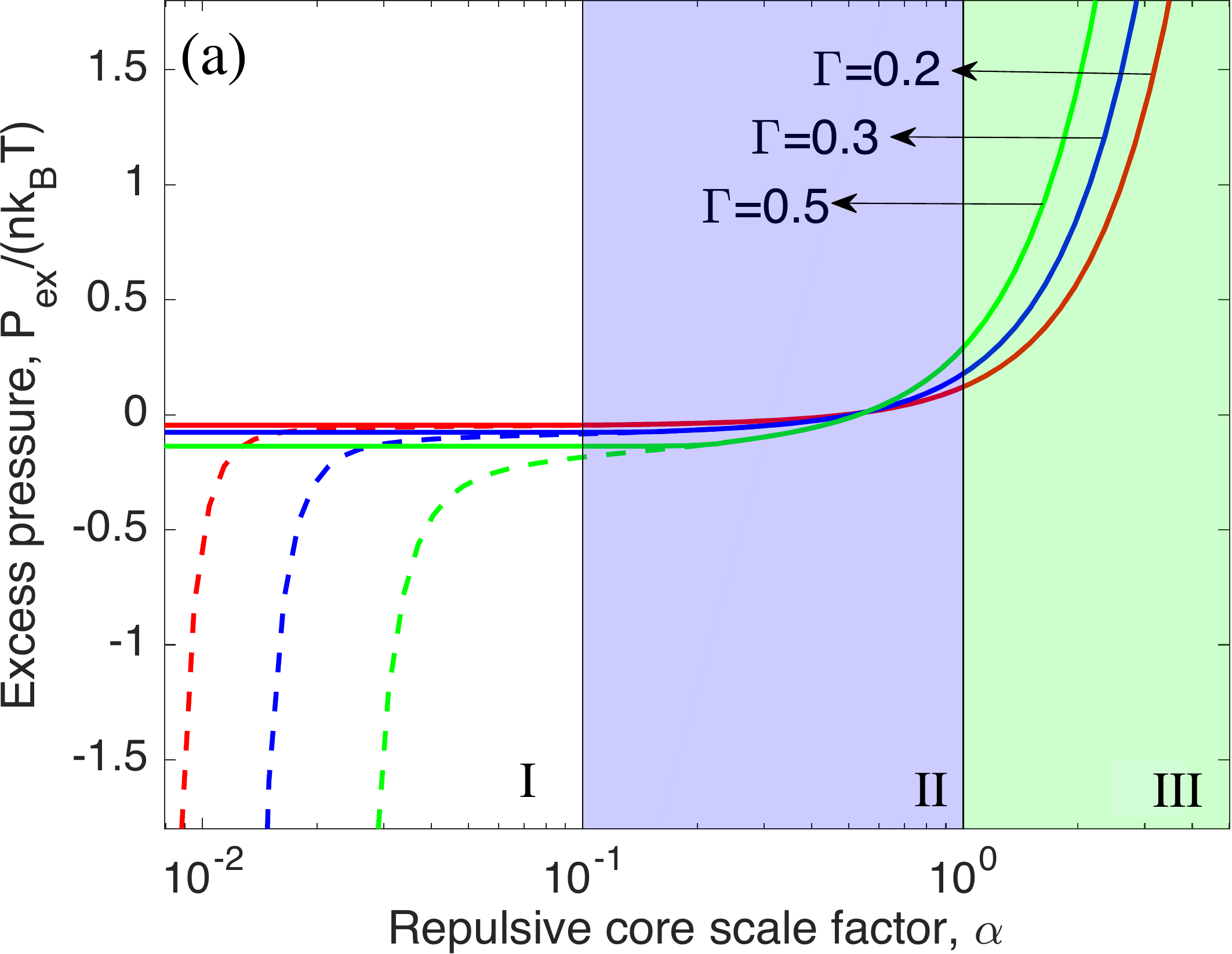}
  \includegraphics[width = 7.0cm]{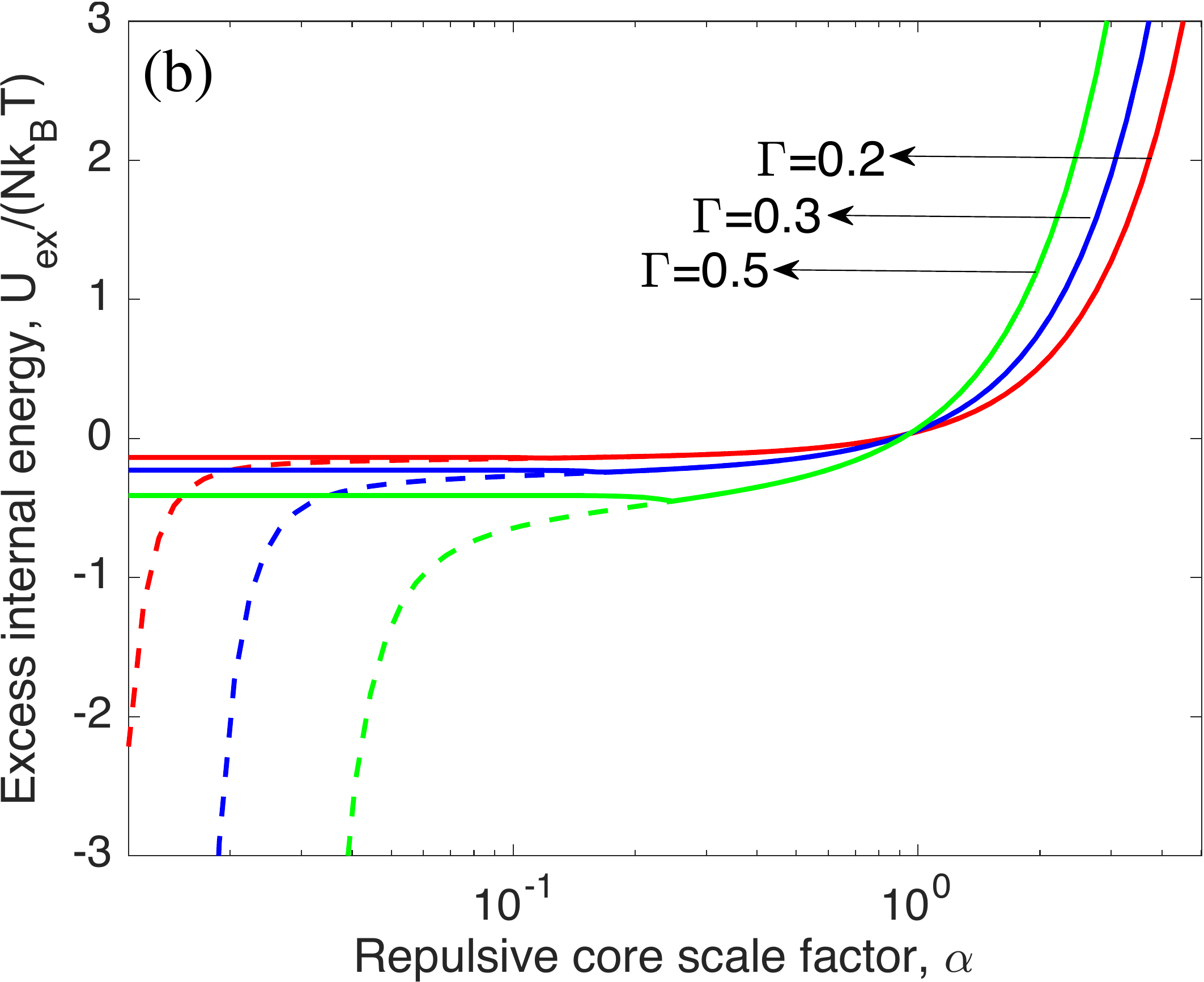}
 \caption{(a) Excess pressure for weakly coupled conditions calculated from Eqs.~\eqref{eq:pots}, \eqref{mean_pot}, and \eqref{pex_we}.  (b) Excess internal energy variation with $\alpha$ and $\Gamma$ using Eq.~(\ref{mean_pot}). In each, dashed lines are for the combined free-plus-bound system. Solid lines contain just the free-charge contribution.} 
 \label{Pex_qu}
\end{figure}

Figure~\ref{Pex_qu}a shows how the excess pressure $P_\textrm{ex}$ varies with $\alpha$ for three values of $\Gamma$.
Based on these curves, we identify three parametric regions.
In the rightmost region III, the repulsive core scale length is larger than the average particle separation ($\alpha \ge 1$). 
Here, the long-range nature of the repulsive cores generates a significant positive excess pressure.   
A physical example of this regime is dense degenerate plasmas where the de Broglie wavelength exceeds the average interparticle spacing. 
In the leftmost region I, the repulsive core scale length scale is much smaller than the average particle spacing ($\alpha \ll 1$). 
Here, the electron-ion potential well is very deep, leading to significant recombination and a corresponding negative excess pressure. 
This is the region of interest for modeling ultracold neutral plasmas. 
In the intermediate region II, the excess pressure takes a constant value that is slightly negative but larger than $-1$, indicating that the total pressure remains positive in this regime. 
\begin{figure}
 \includegraphics[width = 7.0cm]{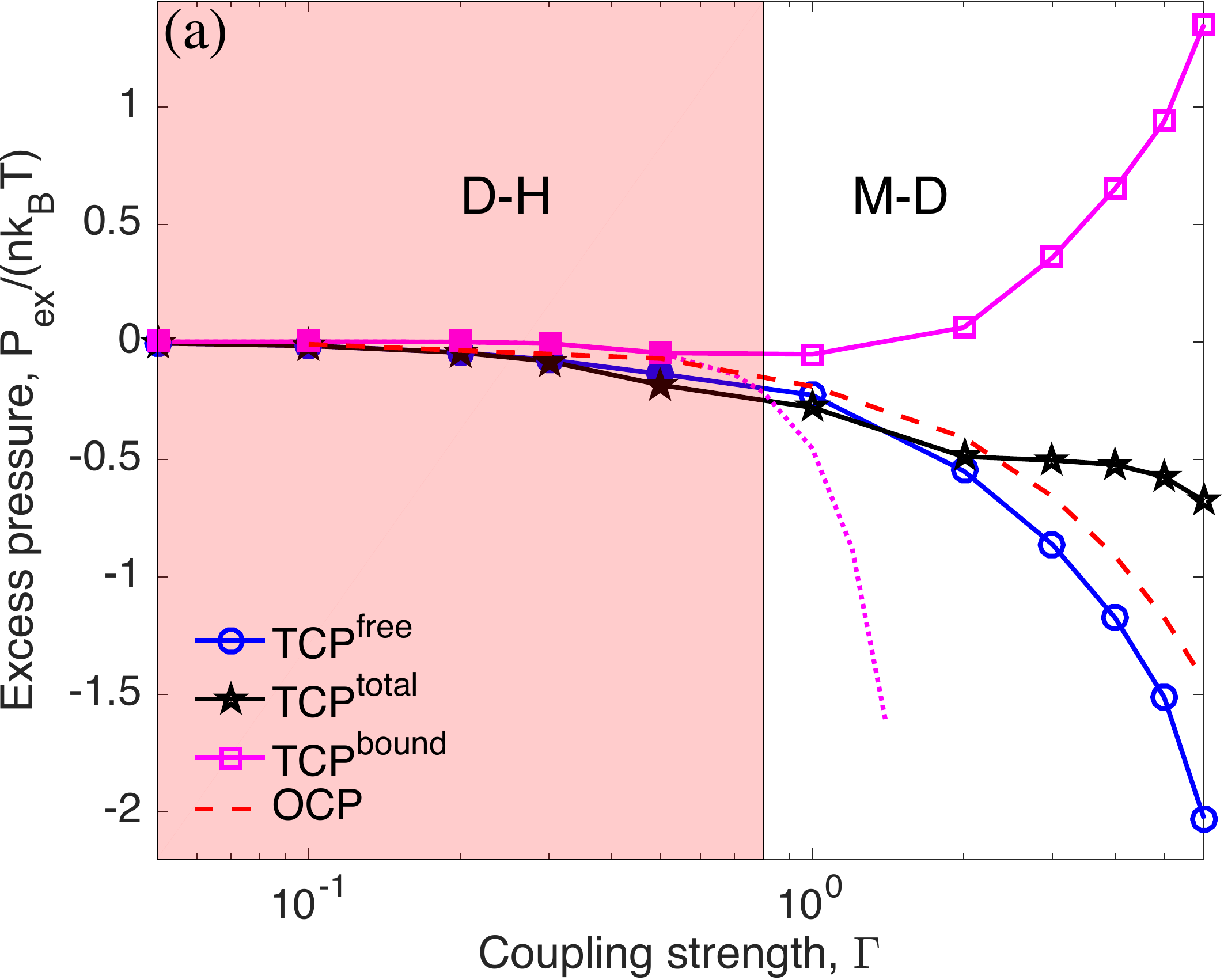}
 \includegraphics[width = 7.0cm]{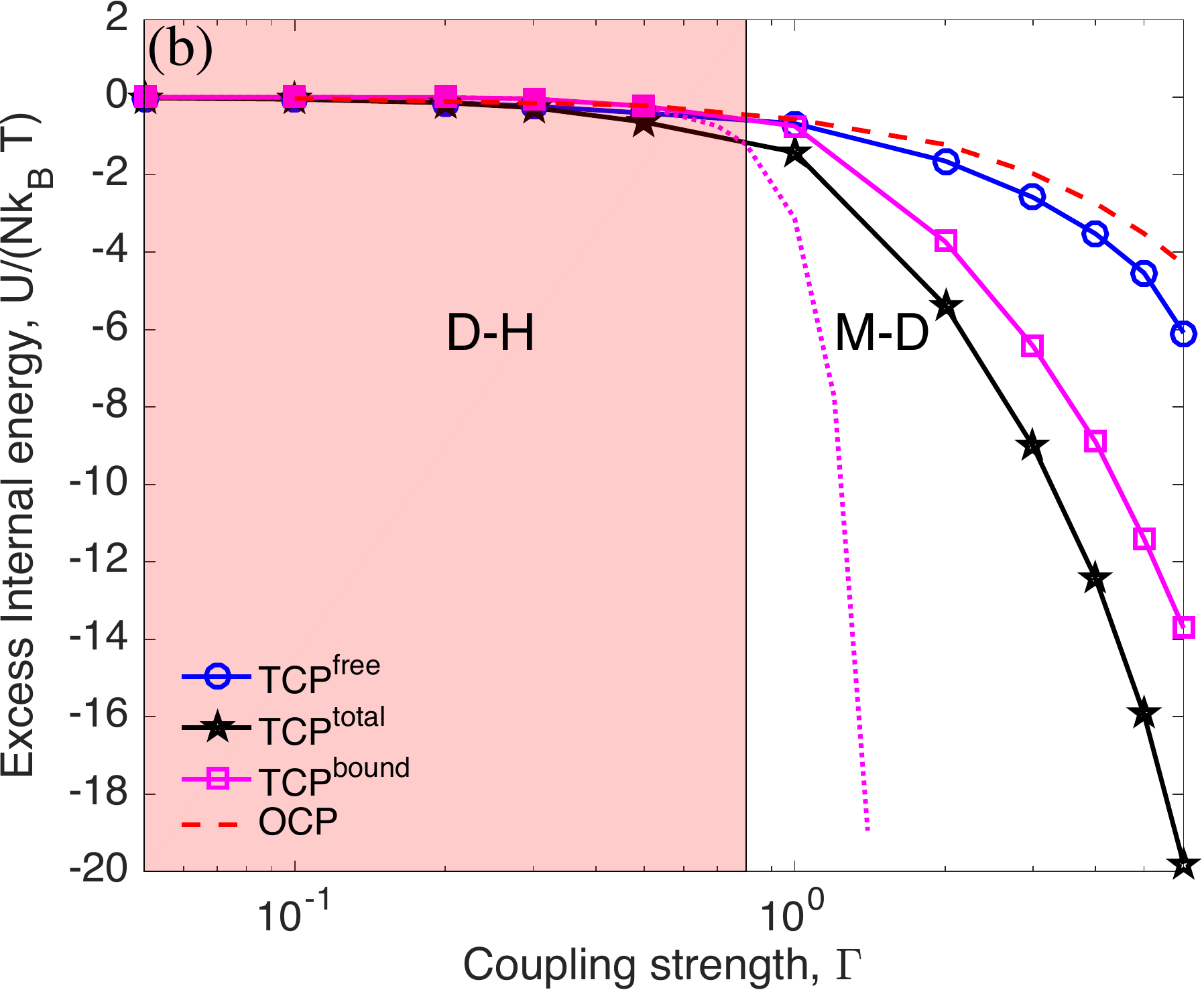}
 \caption{
Dependence on $\Gamma$ of (a) the excess pressure and (b) excess internal energy at $\alpha=0.1$.   Black lines with stars are for the total (bound plus free) system, blue lines with circles are just the free charge contributions, and the magenta line with squares indicates the bound state contributions only. Red dashed lines are the OCP values. Shaded regions indicate where the thermodynamic state variables are calculated using RDFs obtained from Debye-H\"uckel description, while unshaded regions contain molecular dynamics results. Dotted lines represent an extension of Debye-H\"uckel theory into the strongly coupled regime.}
 \label{int_eng}
\end{figure}

The solid lines in Fig.~\ref{Pex_qu}a show the contribution to the excess pressure associated with free charges, which was obtained using the energy criterion in Eq.~\eqref{eq:g_max}. 
Although the excess pressure diverges toward large negative values when the bound states are kept (dashed lines), it is found to be independent of the repulsive core scale parameter $\alpha$ when they are removed (solid lines). 
The asymptotic value associated with the free charge population corresponds to that of the intermediate region II. 
This asymptotic value is what we associate as the excess pressure of the free charge population. 

The $\alpha a$ value separating this intermediate region from region I is associated with the spatial location where the potential energy of the attractive Coulomb interaction significantly exceeds the average kinetic energy. 

In our model, $\alpha$ is a set parameter that is not associated with a physical scale. However, consider for a moment associating the thermal de Broglie wavelength with the repulsive core scale length. The ratio of the thermal de Broglie wavelength and the interparticle spacing $\lambda_{\textrm{db}}/a = [2\pi \hbar^2/(m_{ie} k_BT)]^{1/2}/a$ is a measure of the influence of quantum mechanical wave effects of the ion fluid. Here, $m_{ie} = m_e m_i/(m_e+m_i) \simeq m_e$ is the reduced mass. Applying $\alpha a = \lambda_{\textrm{db}}$, provides $\Gamma a/\lambda_{\textrm{db}} \simeq 2/\sqrt{T[\textrm{eV}]}$. 
Thus, the boundary $\alpha = \lambda_{\textrm{db}}/a \simeq 0.05 \Gamma$ is simply associated with the temperature $T \simeq 0.01$ eV. 
If $T \lesssim 0.01$ eV, the plasma is in region I and the excess pressure is highly negative, indicating the system will collapse (i.e., recombine). 
If $T \gtrsim 0.01$ eV, the plasma is in the plateau region II with a small negative excess pressure, but a positive total pressure. 
Ultracold plasmas fall deep in region I. 
The additional challenge at strong coupling ($\Gamma \gtrsim 1$) is that the intermediate region becomes narrow, and the Debye-H\"{u}ckel approximation breaks down. 
In this region, we will separate the contributions from free charges (plasma) and classical bound states using the same methods outlined in this section, but apply them to RDFs calculated with MD simulations.

Figure~\ref{int_eng}a shows the excess pressure dependence on $\Gamma$ at a fixed value of $\alpha=0.1$. 
Data in the shaded regions was obtained using the Debye-H\"uckel model and data in the non-shaded regions was obtained using molecular dynamics simulations. 
The black line with pentagram markers shows the total excess pressure including free and bound states. 
The blue line with circles denotes the excess partial pressure of the free charges. 
The magenta line with square markers represents bound state contribution. 
At weak coupling, the excess pressure is small, but grows significant as strong coupling is approached
The role of free and bound contributions to the excess pressure in the strongly coupled regime will be further in Sec.~\ref{ex_pr_one}.

\subsection{Internal energy}
\label{exint_ei}
The same arguments used to describe excess pressure in the previous section can be carried over to describe excess internal energy. 
The excess internal energy for an electron-ion plasma can be written in terms of the RDFs as \cite{hansen_pra_81}
\begin{equation}
\frac{U_{ex}}{N} = \frac{2\pi}{n} \sum_{i,j} n_{i} n_{j} \int_0^{\infty}  g_{ij}(r) \left( v_{ij}(r) -T \frac{\partial}{\partial T} v_{ij}(r) \right) r^2 dr~.
\label{uex_unnrm}
\end{equation}
Here $N$ is the total number of particles in the system of volume $V$ such that $n = N/V$.
Note that in the present context $\alpha$ is constant, so the interaction potentials $v_{ij}$ are independent of temperature and the second term in Eq.~\eqref{uex_unnrm} is zero.
However, in a dense plasma context $\alpha a \simeq \lambda_{\textrm{db}}$, so the interaction potentials depend on temperature and this term would be nonzero.

\begin{figure}
 \includegraphics[height=6cm,width = 8.0cm]{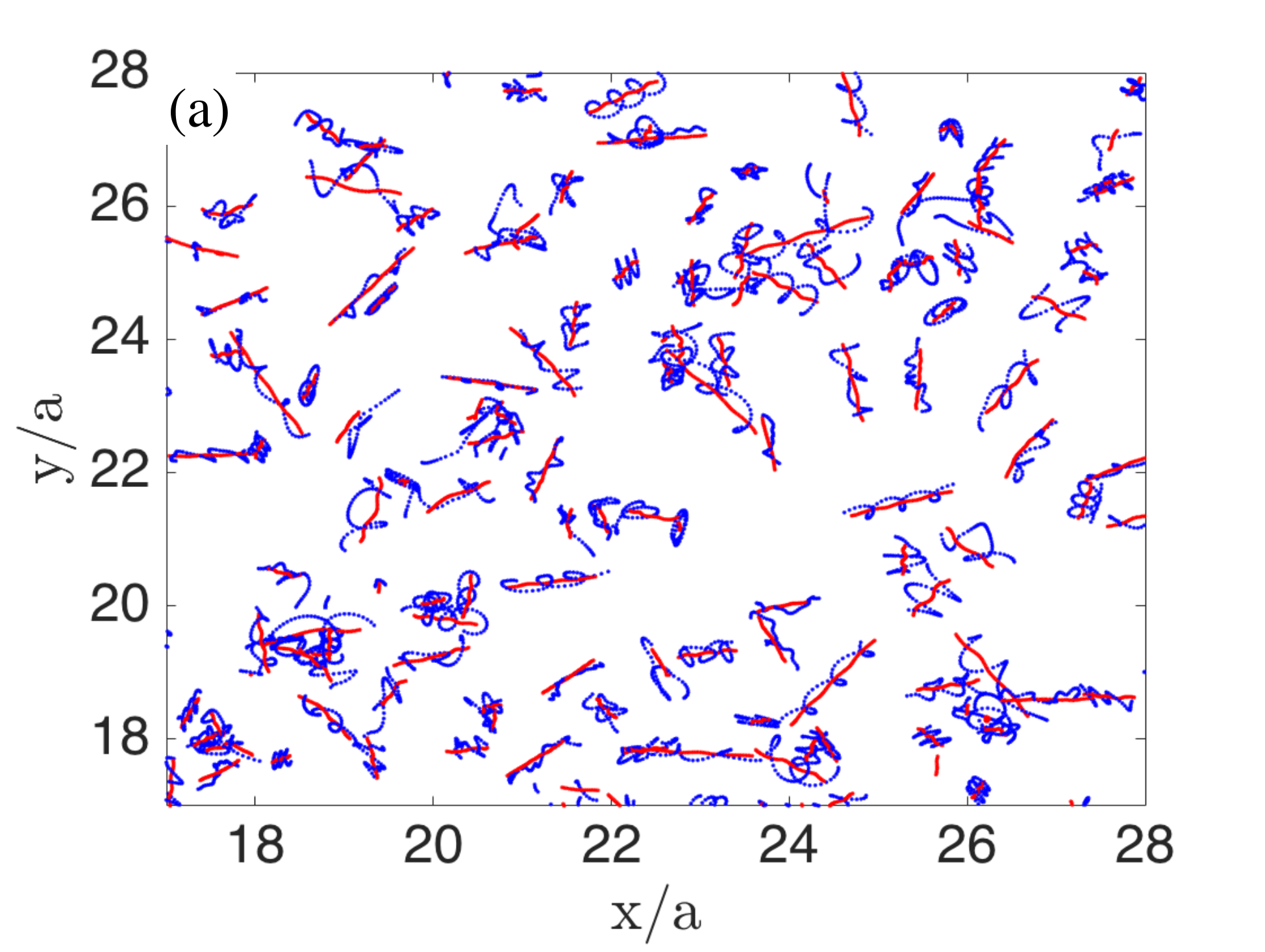}
 \includegraphics[height=6cm,width = 7.0cm]{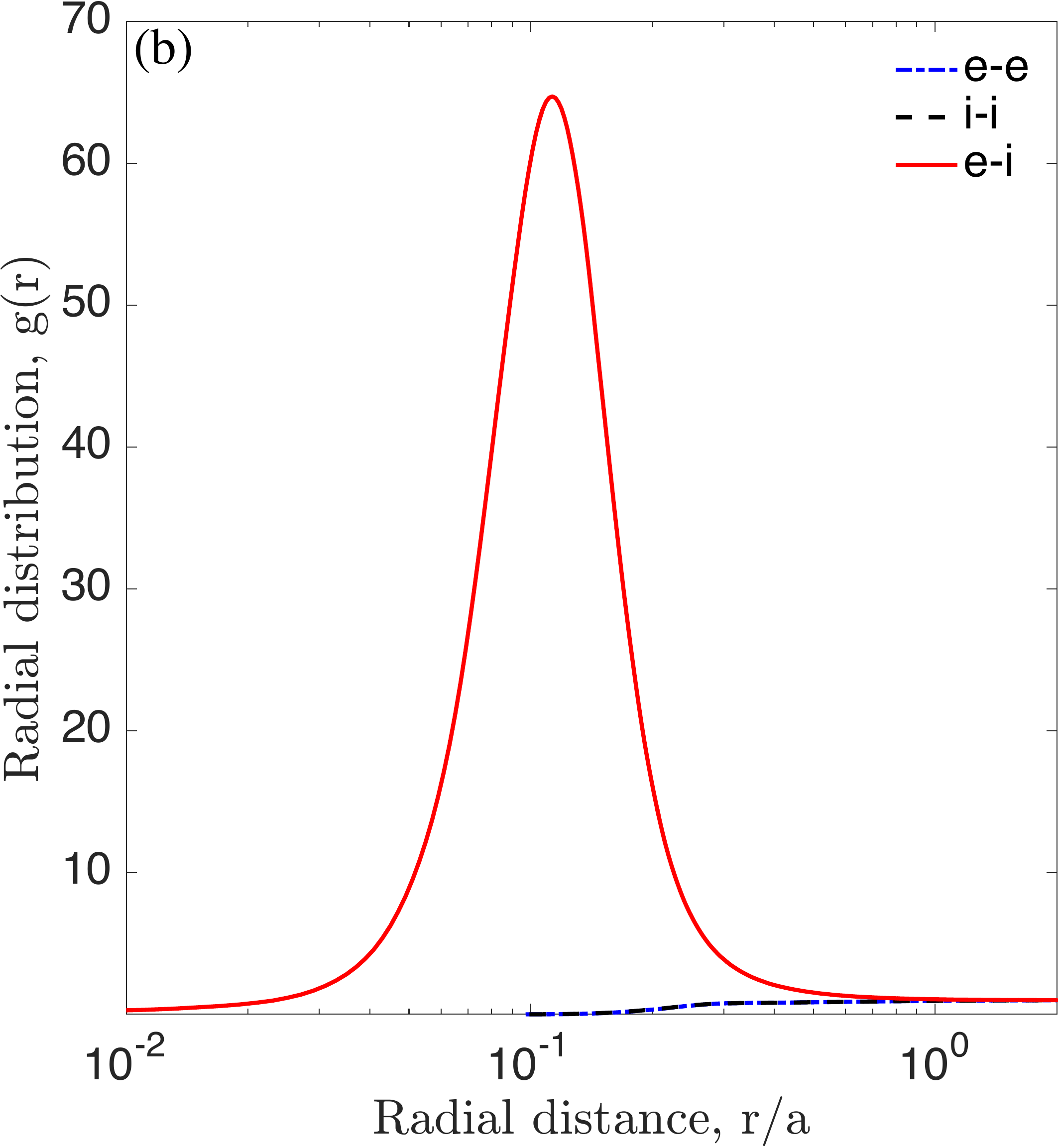}
 \caption{(a) Bound state trajectories of electron-ion pairs during a simulation with $\Gamma = 1$ and $\alpha = 0.1$ over a time interval of $3\omega_{pe}^{-1}$. 
The free particle trajectories have been removed. (b) The RDFs at the same conditions, showing the peak in $g_{ei}(r)$ at $r=\alpha a$.} 
 \label{gr_nd_bound}
\end{figure}

Fig.~\ref{Pex_qu}b shows the variation of internal energy with the repulsive core parameter $\alpha$ for different values of coupling strength $\Gamma$.
As was the case with the excess pressure, the internal energy of the full system (free plus bound) diverges sharply as $\alpha$ decreases, but it asymptotes to a constant when the bound state contribution is removed.

We emphasize that a well-defined thermodynamic pressure and energy for weakly coupled plasmas traditionally rely on being able to neglect the inter-particle interactions in comparison with their kinetic energy. 
The analysis of this section illustrates that the inherent difficulties of a point-particle description of a plasma are still formally present at weak coupling, as evidenced by the negative divergence of the pressure and energy as $\alpha\to{}0$. 
In practice, quantum mechanical effects preventing Coulomb collapse at close distances are responsible for the stability of matter \cite{lieb_76}. 

\section{Simulation model}
\label{sim_mod}
Three dimensional classical MD simulations were carried out using the open source code LAMMPS \cite{Plimpton1995}.
LAMMPS is massively parallel (both CPU and GPU based) and is efficient for large-scale particle simulations. 
The simulation geometry was a 3D cubic box with periodic boundary conditions. 
Each simulation used $10^4$ electrons and $10^4$ ions, and the typical time step was $0.005 \omega_{pe}^{-1}$.
These parameters were chosen to ensure energy conservation as well as to fully resolve the dynamics of the lightest species (i.e. electron)
during the simulation~\cite{dimonte_prl_08}. 
Simulations were conducted by first equilibrating the system using a Nos\'{e}-Hoover thermostat to achieve a desired temperature corresponding to a particular $\Gamma$ value \cite{evans_85}. 
After equilibrium was achieved, the thermostat was turned off and the RDF was computed. 
The PPPM (particle-particle, particle-mesh) method \cite{Plimpton97} was used to calculate the long range interactions.
The interaction potentials used were those from Eq.~\eqref{eq:pots}, with $\alpha$ an input parameter. 
The ion mass was taken to be 10 times higher than the electron mass. 
However, here we present results at equilibrium, in which case we found that the mass ratio did not influence the RDFs, as expected from equilibrium statistical mechanics. 

These  simulations were limited to values of $\alpha$ no less than 0.1 due to energy conservation requirements.
We found that at smaller values of $\alpha$ it became prohibitive to resolve the timescales of tightly bound pairs to the degree required for energy conservation. 
Nevertheless, this value was small enough to reach the desired plateau regime. 

%
\begin{figure}
 \includegraphics[width = 7.0cm]{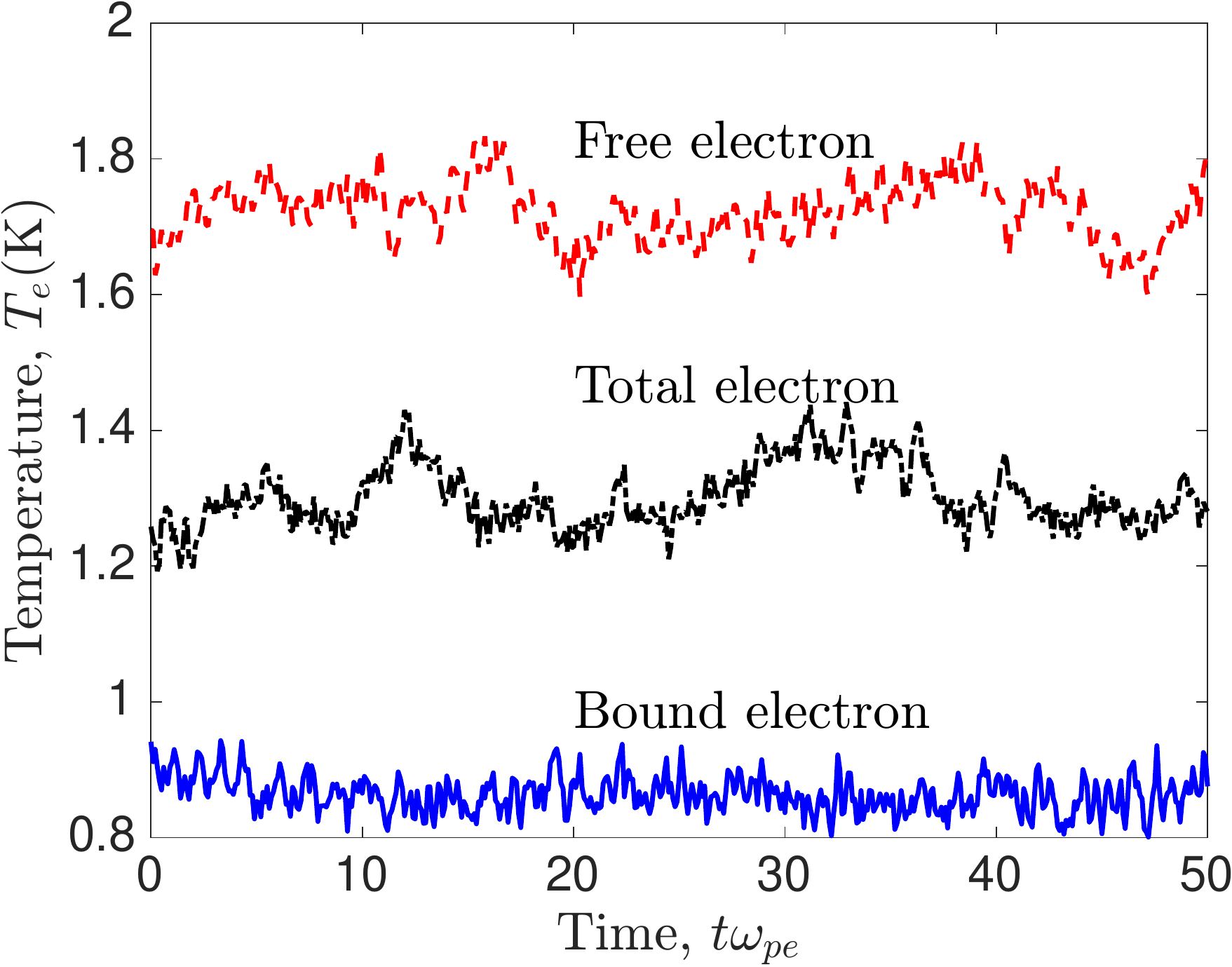}
 \caption{Electron temperature for free (red dashed line) and bound (blue line) species in a simulation with $\Gamma = 1$ and $\alpha = 0.1$. The black line shows the total electron temperature in the system.}
 \label{temp_bound_free}
\end{figure}
%

\begin{figure}
 \includegraphics[width = 8cm]{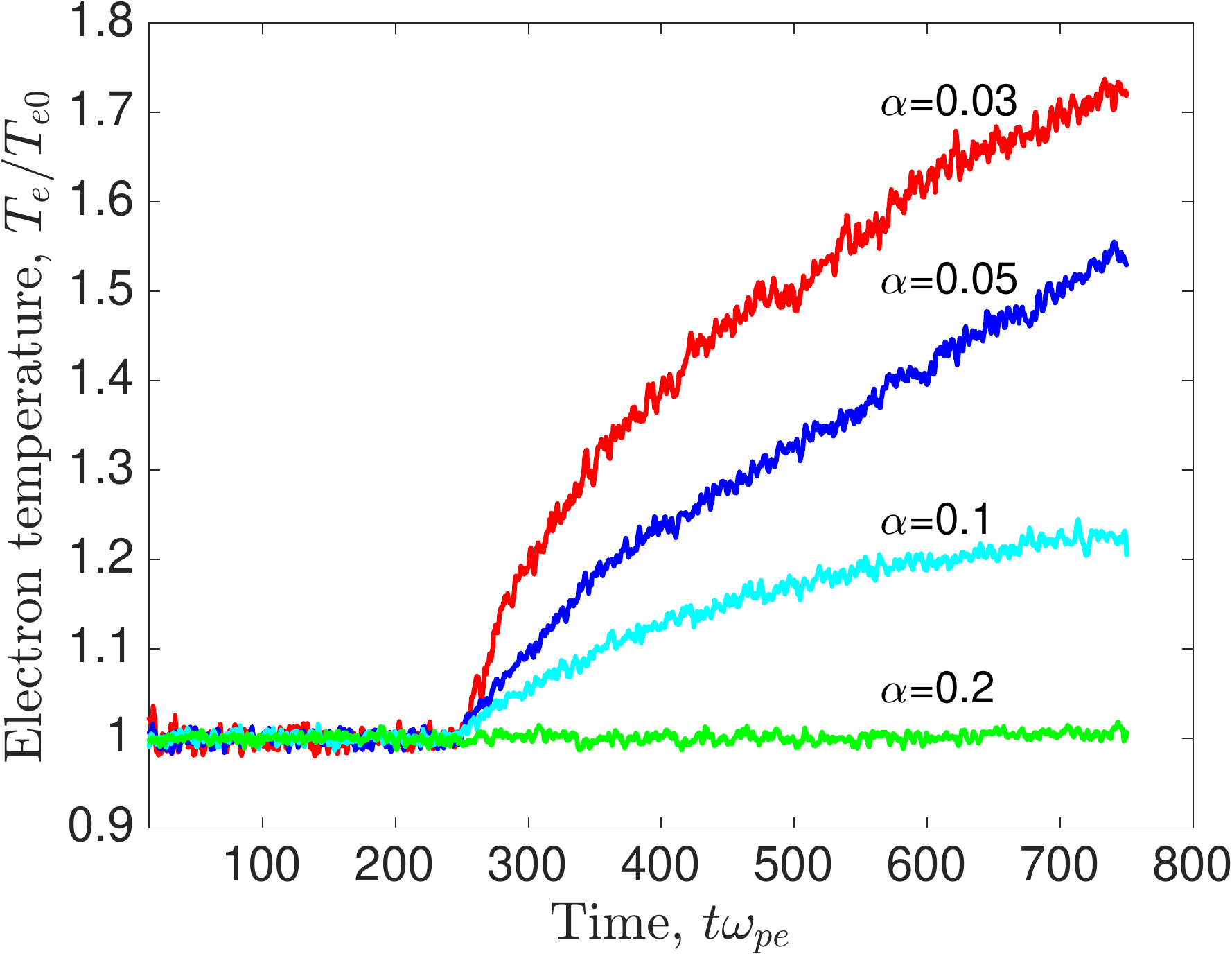}
 \caption{The evolution of the electron temperature in an ultracold plasma simulation at $\Gamma=10$ and various values of $\alpha$. The heating rate increases as the repulsive core distance $\alpha a$ shrinks.
} 
 \label{temp_tcp_evol}
\end{figure}

\section{Strongly coupled plasma}
\label{th_strong}
%
We now apply the concepts and techniques discussed in Section~\ref{th_weak} to moderately and strongly coupled plasmas using classical MD simulations.

\begin{figure*}
 \includegraphics[width = 13cm]{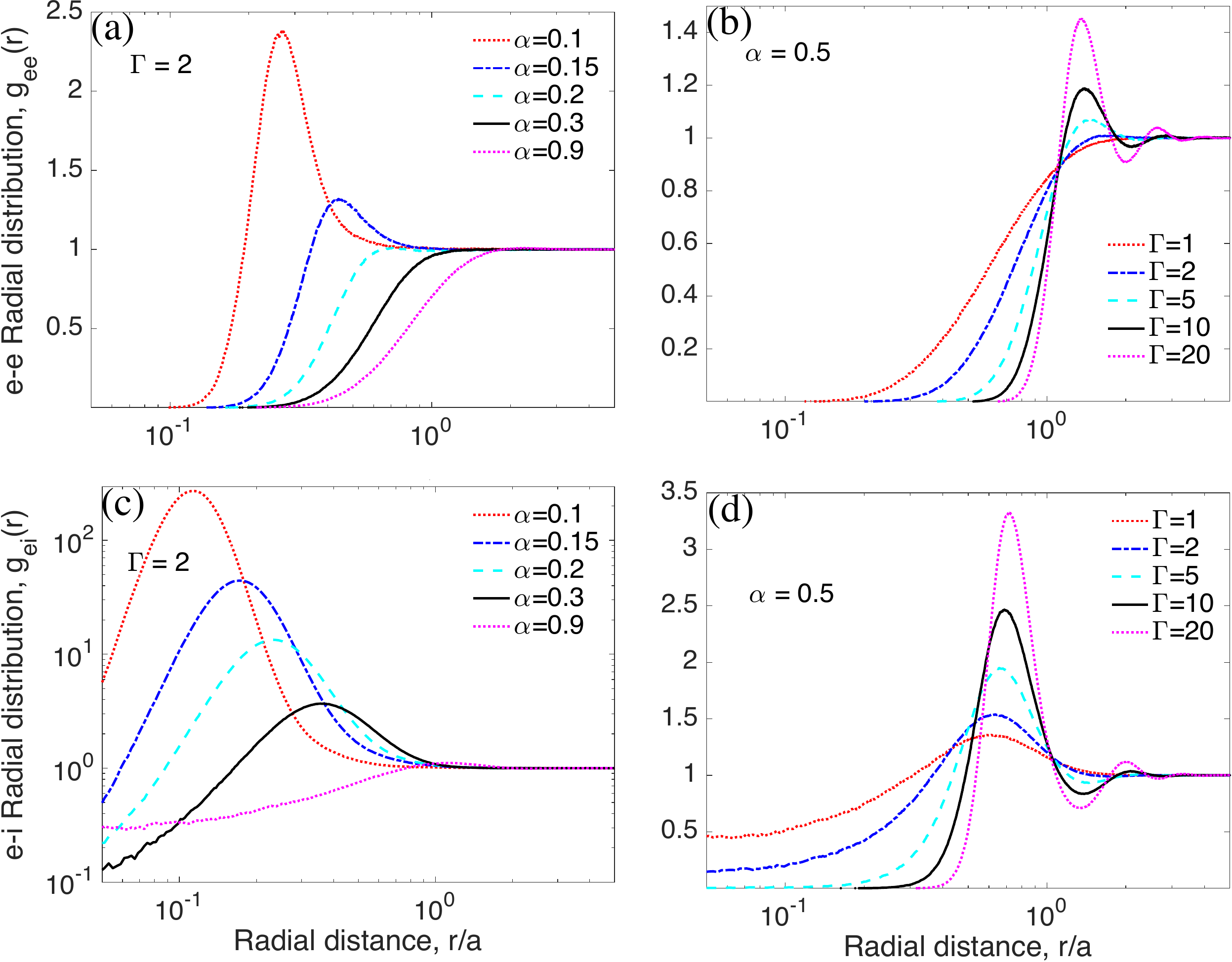}
 \caption{Radial distribution functions $g_{ee}(r)$ (top) and $g_{ei}(r)$ (bottom) for equilibrium electron-ion plasma, showing variation with $\alpha$ (left) and $\Gamma$ (right).}
 \label{gr_th_tcp}
\end{figure*}

\subsection{Classical bound states}
\label{bstate}
Compared to a weakly coupled plasma, a strongly coupled plasma whose particles interact through the pair potentials of Eq.~\eqref{eq:pots} is expected to form more bound states.
This is because the depth of the electron-ion potential well scales linearly with $\Gamma$.
Indeed, comparing Fig.~\ref{gr_nd_bound}b to Fig.~\ref{bd_free}, doubling $\Gamma$ yields nearly a factor of four increase in the peak value of $g_{ei}(r)$ for the same value of $\alpha$, indicating that a larger fraction of the plasma is confined to tight orbits like those pictured in Fig.~\ref{gr_nd_bound}a.
In addition to the many binary bound pairs, we observed that clusters of bound pairs can form stable structures under strong coupling conditions. 
These can take the form of long chains, or rings. 
These structures will be discussed in more detail in a later work. 

In the present studies we limit our results up to moderate coupling strengths only.
The reason is that as we move towards stronger coupling strength, the system's increased affinity for forming bound states results in a ``plasma'' that is primarily composed of clumped bound pairs. 
Removal of the bound states would then effectively take the majority of charged particles out of evaluation of thermodynamic properties, reducing the effective coupling strength of the free charges below their nominal $\Gamma$ value.
Thus our simulation results will not remain practical for higher $\Gamma$ values. 
Physically, this is related to the rapid rate of recombination at these conditions. 

%
\subsection{Three-body interactions}
\label{tbr_sec}
\label{temp_rg}

MD simulations permit us to investigate the cumulative effect of the three-body interactions studied in isolation in Section~\ref{3br_cl}.
To do so, we used a microscopic criterion (instead of Eq.~\eqref{eq:g_max}) to classify individual electron-ion pairs as either free or bound.
For a selected pair of particles, we computed $U_{\mathrm{eff}}(r)$ and $E$ as though the two particles' motion were unaffected by the surrounding plasma.
Repeating for many such pairs, we calculated the kinetic energy for the population of free electrons and bound electrons.
The results for $\Gamma=1$ and $\alpha=0.1$ are plotted in Figure~\ref{temp_bound_free}.
This shows that free electrons carry approximately twice as much kinetic energy than those bound to an ion for this set of parameters.
This is in agreement with our expectations from the three-body dynamics described in Section~\ref{3br_cl}, where electrons that end up in a bound state were observed to give up kinetic energy to other nearby electrons via scattering. 
Figure~\ref{temp_bound_free} also shows that the two electron populations' temperatures remain fixed (aside from fluctuations).
This indicates that not only is the system as a whole in equilibrium, but the free and bound electron sub-systems have each attained their own thermal equilibrium.

We also find that if the thermostat is lifted, the plasma will heat; as evidenced by the temperature evolution plots in Figure~\ref{temp_tcp_evol}.
It can be seen that when the thermostat is switched off at time $t\approx 250\omega_{pe}^{-1}$, the electron temperature increases rapidly if $\alpha$ is sufficiently small. 
Only the electron temperature is shown because the ion temperature curves are identical. 
The heating rate increases as $\alpha$ decreases, implying that the heating of the system arises from the liberation of Coulomb potential energy via classical three-body recombination.
At sufficiently large values of $\alpha$, the heating effect is insignificant even after removing the thermostat, as in the $\alpha=0.2$ line of Figure~\ref{temp_tcp_evol}.
However, for smaller values of $\alpha$, the heating effects become more and more significant because the deeper potential well in the electron-ion interaction provides a larger potential energy source that is converted to kinetic energy via heating. 
This figure demonstrates the rapid evolution of ultracold plasmas, and that the concept of quasi-equilibrium relates to a narrow time window. 
In contrast, the same concept is much more clearly defined in dense degenerate plasmas with slow recombination rates \cite{hansen_pra_81,glosli_pre_08}.

\subsection{Radial distribution functions}
\label{rdfs}

Figure~\ref{gr_th_tcp} shows the RDFs for electron-electron and electron-ion pairs obtained from classical MD simulation of an ultracold plasma.
The upper panels show the RDFs for electron-electron pairs while the lower panel shows the RDFs for electron-ion pairs.
Subplots (a) and (c) show the effect of varying $\alpha$ at fixed $\Gamma$, and vice-versa for subplots (b) and (d).

In addition to the peak in $g_{ei}(r)$ near $r=\alpha a$, the figures show an additional peak in $g_{ee/ii}(r)$.
This feature of the like-charge RDFs is a consequence of the system's tendency to cluster.
Tightly bound e-i pairs are essentially dipoles, which attract other dipoles and cause clumping to occur.
This permits, for example, two bound electrons to lie near each other in spite of their mutual repulsion.

Though interesting, this secondary peak in $g_{ee/ii}(r)$ complicates the procedure for separating bound and free populations at the RDF level.
In order to use Eq.~\eqref{eq:g_max} to remove bound states from $g_{ei}(r)$, we must also remove them from $g_{ee/ii}(r)$ in an internally consistent way.
To do so, we first enforce Eq.~\eqref{eq:g_max} as before, yielding a cutoff distance $r_{c,ei}$.
Next, we determine a second cutoff distance $r_{c,ii}(=r_{c,ee})$ that maintains quasineutrality within the individual bound and free subpopulations.
That is, we determine $r_{c,ii/ee}$ such that 
\begin{subequations}
\begin{align}
&4\pi n_i \int_{r_{c,ii}}^{\infty} g_{ii}(r) r^2 dr - 4\pi n_e \int_{r_{c,ei}}^\infty g_{ei}(r) r^2 dr = -1  \\
&4\pi n_i \int_{r_{c,ei}}^{\infty} g_{ei}(r) r^2 dr - 4\pi n_e \int_{r_{c,ee}}^\infty g_{ee}(r) r^2 dr = 1. 
\end{align}
\label{sum_bf}
\end{subequations}
Again, the hypothesis here is that approximately all contributions to $g_{ii}(r)$ below $r_{c,ii}$ are due to clustering of bound pairs.

\subsection{Excess pressure}
\label{ex_pr_one}
%
\begin{figure}
 \includegraphics[width = 7.0cm]{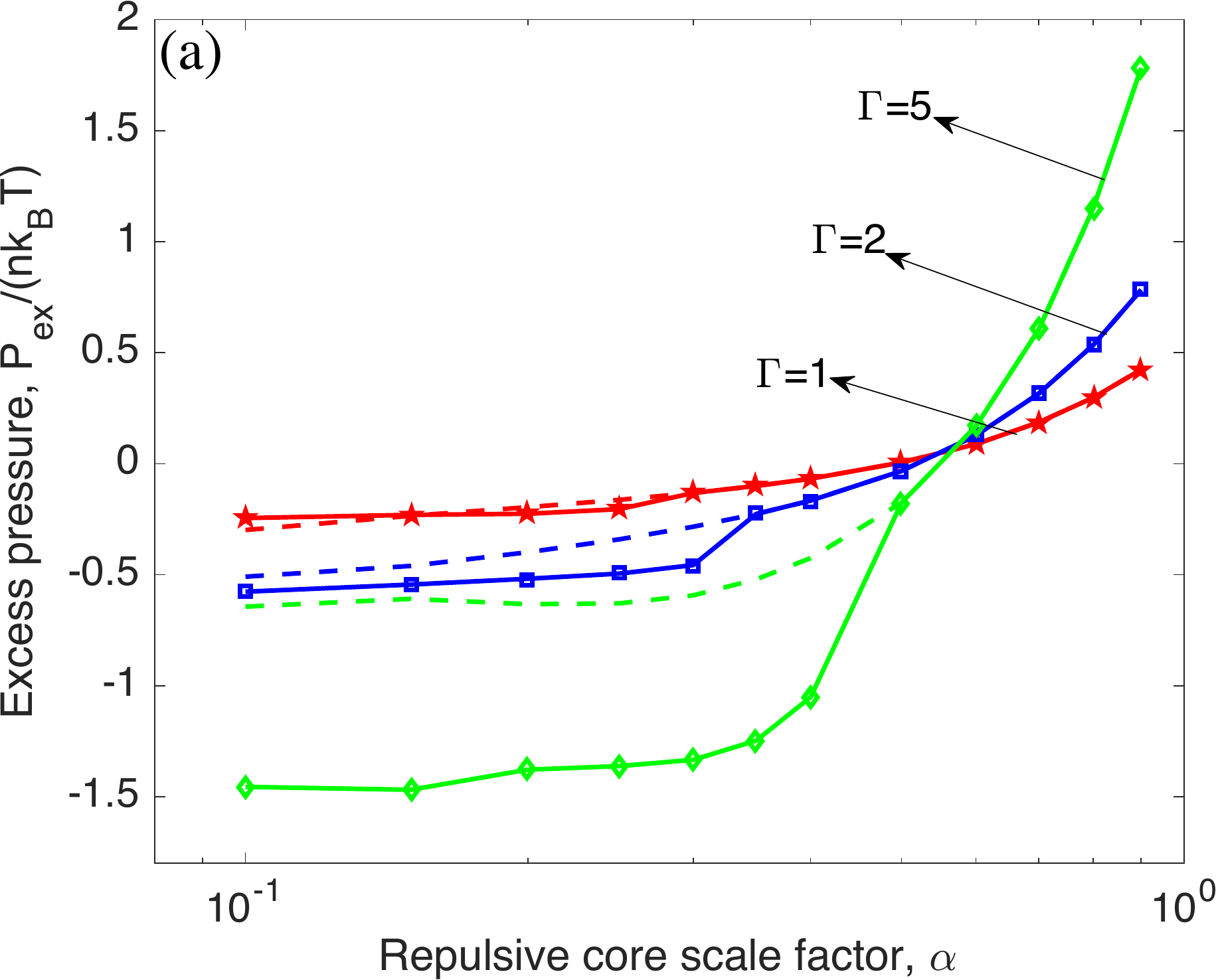}
 \includegraphics[width = 7.0cm]{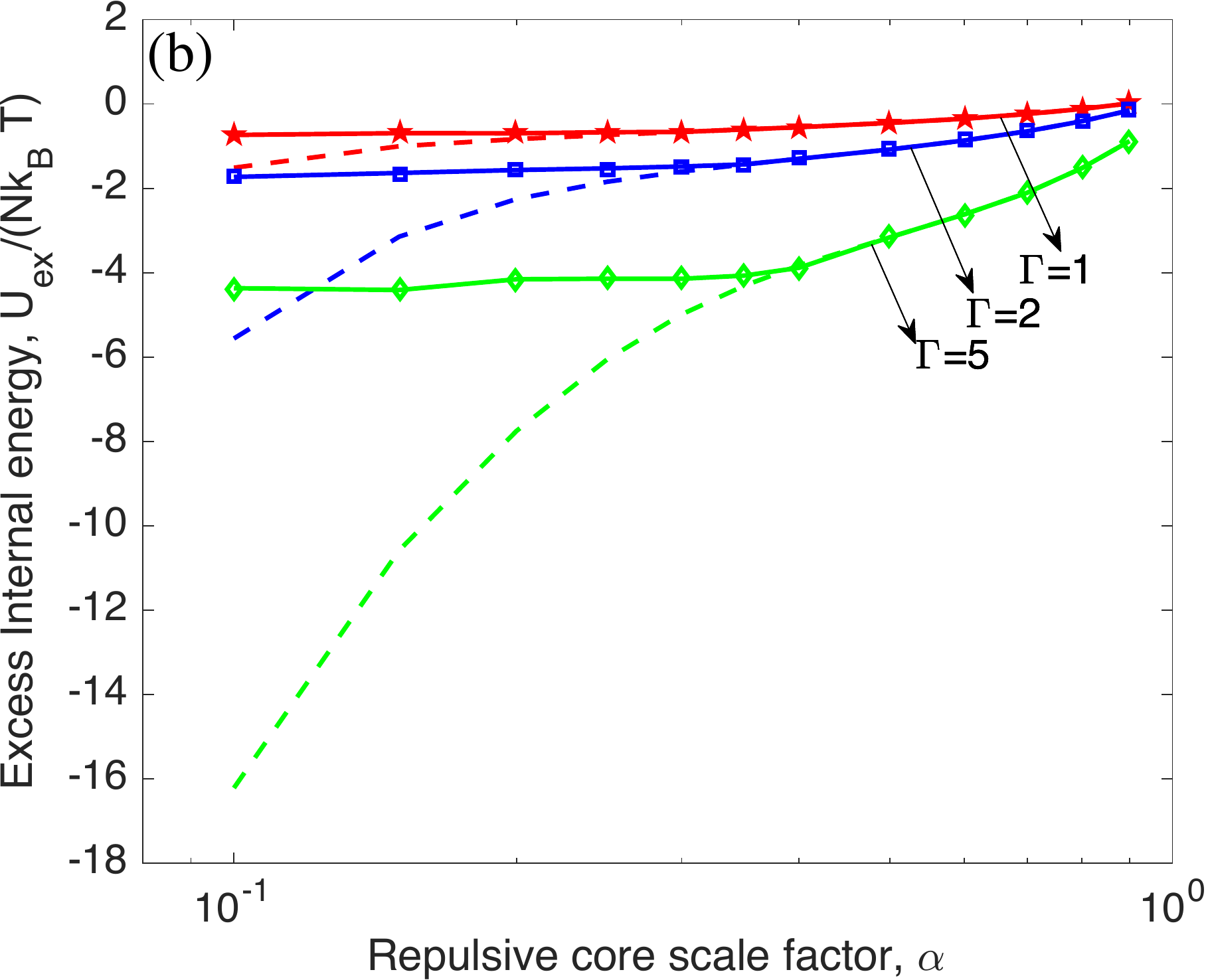}
 \caption{Dependence of the excess pressure (a) and excess internal energy (b) on the repulsive core parameter $\alpha$ at various coupling strengths.  
 Dashed lines indicate results for the combined free-plus-bound system and the solid lines with markers indicated results isolating the free-charge components of the system.}
 \label{Pex_md}
\end{figure}
The excess pressure for moderately coupled ultracold plasmas was evaluated using Eq.~\eqref{pex_we}.
The input RDFs for these moderately coupled media were obtained from equilibrium MD simulations. 
In Fig.~\ref{Pex_md}a, the total excess pressure (including both free and bound charges) is plotted as a function of $\alpha$ for $\Gamma=1,2,5$.
The line plots with markers show the excess pressure calculated after removal of bound states from the RDFs.
The bound- and free-state RDFs have been separated using Eqs.~\eqref{eq:g_max} and~\eqref{sum_bf}.
Like the weakly coupled regime from Fig.~\ref{Pex_qu}, the excess pressure for the free charge population is found to  plateau to an $\alpha$-independent value for small $\alpha$.

Fig.~\ref{int_eng}a shows the excess pressure dependence on $\Gamma$ with $\alpha$ fixed at a value of  $0.1$. 
The pressure of the total plasma remains positive ($P_{ex}>-1$) at all values of $\Gamma$ shown.
The blue line with circles and magenta line with squares are the partial excess pressures due to free and bound charges, respectively.
This figure shows that the partial excess pressure due to the free charge contribution closely follows the OCP results, i.e., $P_{ex}<0$ and that the total partial pressure associated with the free charges becomes negative around $\Gamma \approx 4$. 
The collapsing nature of free electron-ion gas is responsible for this negative excess pressure.
The partial excess pressure due to bound states is always found to be positive, much like what one would expect for a gas of neutral atoms. 
The figure also shows that the MD results show a consistent trend that merges with the Debye-H\"{u}ckel results at weak coupling. 
\subsection{Internal energy}
\label{int_ei}
The excess internal energy was also evaluated using the RDFs obtained from MD along with Eq.~\eqref{uex_unnrm}.
Figure~\ref{Pex_md}b shows how the excess internal energy of the total (free plus bound) plasma depends on $\alpha$ at $\Gamma=1,2,5$.
The lines with markers represent the excess internal energy of free charges found by removing the contribution of bound states to $g_{ij}(r)$.
Similarly to previous results for excess pressure, the removal of bound states again leads to values of $U_{\mathrm{ex}}$ that are independent of  $\alpha$ in the $\alpha\to 0$ limit.
 
 Fig.~\ref{int_eng}b shows the excess internal energy at different values of coupling strength $\Gamma$ with $\alpha= 0.1$.
 Similar to excess pressure, the excess internal energy of the free states (blue line with stars) is close to the OCP value (red dashed line).
At increased coupling strength, an increase in the bound state fraction causes the internal energy to become increasingly negative.

\section{Conclusions}
\label{concl}
Using Debye-H\"{u}ckel theory for the weakly coupled regime and equilibrium MD simulations for the strongly coupled regime, we observed that the Coulomb collapse of a classical electron-ion plasma can be prevented by applying a repulsive core force at close distance ($\alpha a$) in the electron-ion interaction.
Furthermore, the removal of the bound state contribution to the radial distribution functions was shown to provide predictions for the thermodynamic state variables that are independent of the model repulsive core length scale.

These results provide a method to separate the contribution of free charges from classical bound states in the evaluation of pressure and internal energy of a classical electron-ion plasma. 
This enables quasi-equilibrium analysis of classical electron-ion plasmas, as are found in ultracold neutral plasma experiments. 
Such an analysis is useful for connecting theoretical predictions, which are made at fixed conditions, with experimental measurements, which are made over short enough time intervals that the conditions may be considered fixed.  
The work lays important groundwork for the further development of two-component models for ultracold plasmas based on a classical point-particle picture of the microscopic dynamics.
This study was limited to moderate coupling strengths due to the formation of complex bound-state structures at higher coupling strengths.
Future studies will investigate these structures in further detail.


\begin{acknowledgments}

This material is based upon work supported by the National Science Foundation under Grant No.~PHY-1453736. 
It used the Extreme Science and Engineering Discovery Environment (XSEDE), which is supported by NSF grant number ACI-1053575, under project award No.~PHY-150018.

\end{acknowledgments}

\bibliography{thermo_ei}

\begin{thebibliography}{33}%
\makeatletter
\providecommand \@ifxundefined [1]{%
 \@ifx{#1\undefined}
}%
\providecommand \@ifnum [1]{%
 \ifnum #1\expandafter \@firstoftwo
 \else \expandafter \@secondoftwo
 \fi
}%
\providecommand \@ifx [1]{%
 \ifx #1\expandafter \@firstoftwo
 \else \expandafter \@secondoftwo
 \fi
}%
\providecommand \natexlab [1]{#1}%
\providecommand \enquote  [1]{``#1''}%
\providecommand \bibnamefont  [1]{#1}%
\providecommand \bibfnamefont [1]{#1}%
\providecommand \citenamefont [1]{#1}%
\providecommand \href@noop [0]{\@secondoftwo}%
\providecommand \href [0]{\begingroup \@sanitize@url \@href}%
\providecommand \@href[1]{\@@startlink{#1}\@@href}%
\providecommand \@@href[1]{\endgroup#1\@@endlink}%
\providecommand \@sanitize@url [0]{\catcode `\\12\catcode `\$12\catcode
  `\&12\catcode `\#12\catcode `\^12\catcode `\_12\catcode `\%12\relax}%
\providecommand \@@startlink[1]{}%
\providecommand \@@endlink[0]{}%
\providecommand \url  [0]{\begingroup\@sanitize@url \@url }%
\providecommand \@url [1]{\endgroup\@href {#1}{\urlprefix }}%
\providecommand \urlprefix  [0]{URL }%
\providecommand \Eprint [0]{\href }%
\providecommand \doibase [0]{http://dx.doi.org/}%
\providecommand \selectlanguage [0]{\@gobble}%
\providecommand \bibinfo  [0]{\@secondoftwo}%
\providecommand \bibfield  [0]{\@secondoftwo}%
\providecommand \translation [1]{[#1]}%
\providecommand \BibitemOpen [0]{}%
\providecommand \bibitemStop [0]{}%
\providecommand \bibitemNoStop [0]{.\EOS\space}%
\providecommand \EOS [0]{\spacefactor3000\relax}%
\providecommand \BibitemShut  [1]{\csname bibitem#1\endcsname}%
\let\auto@bib@innerbib\@empty
\bibitem [{\citenamefont {Diaw}\ and\ \citenamefont {Murillo}(2015)}]{diaw_15}%
  \BibitemOpen
  \bibfield  {author} {\bibinfo {author} {\bibfnamefont {A.}~\bibnamefont
  {Diaw}}\ and\ \bibinfo {author} {\bibfnamefont {M.~S.}\ \bibnamefont
  {Murillo}},\ }\href {\doibase 10.1103/PhysRevE.92.013107} {\bibfield
  {journal} {\bibinfo  {journal} {Phys. Rev. E}\ }\textbf {\bibinfo {volume}
  {92}},\ \bibinfo {pages} {013107} (\bibinfo {year} {2015})}\BibitemShut
  {NoStop}%
\bibitem [{\citenamefont {Strickler}\ \emph {et~al.}(2016)\citenamefont
  {Strickler}, \citenamefont {Langin}, \citenamefont {McQuillen}, \citenamefont
  {Daligault},\ and\ \citenamefont {Killian}}]{strickler_16}%
  \BibitemOpen
  \bibfield  {author} {\bibinfo {author} {\bibfnamefont {T.~S.}\ \bibnamefont
  {Strickler}}, \bibinfo {author} {\bibfnamefont {T.~K.}\ \bibnamefont
  {Langin}}, \bibinfo {author} {\bibfnamefont {P.}~\bibnamefont {McQuillen}},
  \bibinfo {author} {\bibfnamefont {J.}~\bibnamefont {Daligault}}, \ and\
  \bibinfo {author} {\bibfnamefont {T.~C.}\ \bibnamefont {Killian}},\ }\href
  {\doibase 10.1103/PhysRevX.6.021021} {\bibfield  {journal} {\bibinfo
  {journal} {Phys. Rev. X}\ }\textbf {\bibinfo {volume} {6}},\ \bibinfo {pages}
  {021021} (\bibinfo {year} {2016})}\BibitemShut {NoStop}%
\bibitem [{\citenamefont {Drake}(2006)}]{drake2006high}%
  \BibitemOpen
  \bibfield  {author} {\bibinfo {author} {\bibfnamefont {R.}~\bibnamefont
  {Drake}},\ }\href {https://books.google.com/books?id=HBj2gqoAy0sC} {\emph
  {\bibinfo {title} {High-Energy-Density Physics: Fundamentals, Inertial
  Fusion, and Experimental Astrophysics}}},\ Shock Wave and High Pressure
  Phenomena\ (\bibinfo  {publisher} {Springer Berlin Heidelberg},\ \bibinfo
  {year} {2006})\BibitemShut {NoStop}%
\bibitem [{\citenamefont {Redmer}\ and\ \citenamefont
  {R\"opke}(2010)}]{redmer_10}%
  \BibitemOpen
  \bibfield  {author} {\bibinfo {author} {\bibfnamefont {R.}~\bibnamefont
  {Redmer}}\ and\ \bibinfo {author} {\bibfnamefont {G.}~\bibnamefont
  {R\"opke}},\ }\href {\doibase 10.1002/ctpp.201000079} {\bibfield  {journal}
  {\bibinfo  {journal} {Contributions to Plasma Physics}\ }\textbf {\bibinfo
  {volume} {50}},\ \bibinfo {pages} {970} (\bibinfo {year} {2010})}\BibitemShut
  {NoStop}%
\bibitem [{\citenamefont {Brueckner}\ and\ \citenamefont
  {Jorna}(1974)}]{Brueckner_74}%
  \BibitemOpen
  \bibfield  {author} {\bibinfo {author} {\bibfnamefont {K.~A.}\ \bibnamefont
  {Brueckner}}\ and\ \bibinfo {author} {\bibfnamefont {S.}~\bibnamefont
  {Jorna}},\ }\href {\doibase 10.1103/RevModPhys.46.325} {\bibfield  {journal}
  {\bibinfo  {journal} {Rev. Mod. Phys.}\ }\textbf {\bibinfo {volume} {46}},\
  \bibinfo {pages} {325} (\bibinfo {year} {1974})}\BibitemShut {NoStop}%
\bibitem [{\citenamefont {Baus}\ and\ \citenamefont
  {Hansen}(1980)}]{Baus19801}%
  \BibitemOpen
  \bibfield  {author} {\bibinfo {author} {\bibfnamefont {M.}~\bibnamefont
  {Baus}}\ and\ \bibinfo {author} {\bibfnamefont {J.-P.}\ \bibnamefont
  {Hansen}},\ }\href {\doibase http://dx.doi.org/10.1016/0370-1573(80)90022-8}
  {\bibfield  {journal} {\bibinfo  {journal} {Physics Reports}\ }\textbf
  {\bibinfo {volume} {59}},\ \bibinfo {pages} {1 } (\bibinfo {year}
  {1980})}\BibitemShut {NoStop}%
\bibitem [{\citenamefont {Baus}(1978)}]{baus_78}%
  \BibitemOpen
  \bibfield  {author} {\bibinfo {author} {\bibfnamefont {M.}~\bibnamefont
  {Baus}},\ }\href {http://stacks.iop.org/0305-4470/11/i=12/a=011} {\bibfield
  {journal} {\bibinfo  {journal} {Journal of Physics A: Mathematical and
  General}\ }\textbf {\bibinfo {volume} {11}},\ \bibinfo {pages} {2451}
  (\bibinfo {year} {1978})}\BibitemShut {NoStop}%
\bibitem [{\citenamefont {Bonitz}\ \emph {et~al.}(2004)\citenamefont {Bonitz},
  \citenamefont {Zelener}, \citenamefont {Zelener}, \citenamefont {Manykin},
  \citenamefont {Filinov},\ and\ \citenamefont {Fortov}}]{Bonitz_04}%
  \BibitemOpen
  \bibfield  {author} {\bibinfo {author} {\bibfnamefont {M.}~\bibnamefont
  {Bonitz}}, \bibinfo {author} {\bibfnamefont {B.~B.}\ \bibnamefont {Zelener}},
  \bibinfo {author} {\bibfnamefont {B.~V.}\ \bibnamefont {Zelener}}, \bibinfo
  {author} {\bibfnamefont {E.~A.}\ \bibnamefont {Manykin}}, \bibinfo {author}
  {\bibfnamefont {V.~S.}\ \bibnamefont {Filinov}}, \ and\ \bibinfo {author}
  {\bibfnamefont {V.~E.}\ \bibnamefont {Fortov}},\ }\href {\doibase
  10.1134/1.1757672} {\bibfield  {journal} {\bibinfo  {journal} {Journal of
  Experimental and Theoretical Physics}\ }\textbf {\bibinfo {volume} {98}},\
  \bibinfo {pages} {719} (\bibinfo {year} {2004})}\BibitemShut {NoStop}%
\bibitem [{\citenamefont {Killian}\ \emph {et~al.}(1999)\citenamefont
  {Killian}, \citenamefont {Kulin}, \citenamefont {Bergeson}, \citenamefont
  {Orozco}, \citenamefont {Orzel},\ and\ \citenamefont
  {Rolston}}]{killian_prl_99}%
  \BibitemOpen
  \bibfield  {author} {\bibinfo {author} {\bibfnamefont {T.~C.}\ \bibnamefont
  {Killian}}, \bibinfo {author} {\bibfnamefont {S.}~\bibnamefont {Kulin}},
  \bibinfo {author} {\bibfnamefont {S.~D.}\ \bibnamefont {Bergeson}}, \bibinfo
  {author} {\bibfnamefont {L.~A.}\ \bibnamefont {Orozco}}, \bibinfo {author}
  {\bibfnamefont {C.}~\bibnamefont {Orzel}}, \ and\ \bibinfo {author}
  {\bibfnamefont {S.~L.}\ \bibnamefont {Rolston}},\ }\href {\doibase
  10.1103/PhysRevLett.83.4776} {\bibfield  {journal} {\bibinfo  {journal}
  {Phys. Rev. Lett.}\ }\textbf {\bibinfo {volume} {83}},\ \bibinfo {pages}
  {4776} (\bibinfo {year} {1999})}\BibitemShut {NoStop}%
\bibitem [{\citenamefont {Killian}(2007)}]{Killian705}%
  \BibitemOpen
  \bibfield  {author} {\bibinfo {author} {\bibfnamefont {T.~C.}\ \bibnamefont
  {Killian}},\ }\href {\doibase 10.1126/science.1130556} {\bibfield  {journal}
  {\bibinfo  {journal} {Science}\ }\textbf {\bibinfo {volume} {316}},\ \bibinfo
  {pages} {705} (\bibinfo {year} {2007})}\BibitemShut {NoStop}%
\bibitem [{\citenamefont {Killian}\ \emph {et~al.}(2007)\citenamefont
  {Killian}, \citenamefont {Pattard}, \citenamefont {Pohl},\ and\ \citenamefont
  {Rost}}]{Killian2007}%
  \BibitemOpen
  \bibfield  {author} {\bibinfo {author} {\bibfnamefont {T.}~\bibnamefont
  {Killian}}, \bibinfo {author} {\bibfnamefont {T.}~\bibnamefont {Pattard}},
  \bibinfo {author} {\bibfnamefont {T.}~\bibnamefont {Pohl}}, \ and\ \bibinfo
  {author} {\bibfnamefont {J.}~\bibnamefont {Rost}},\ }\href {\doibase
  http://dx.doi.org/10.1016/j.physrep.2007.04.007} {\bibfield  {journal}
  {\bibinfo  {journal} {Physics Reports}\ }\textbf {\bibinfo {volume} {449}},\
  \bibinfo {pages} {77 } (\bibinfo {year} {2007})}\BibitemShut {NoStop}%
\bibitem [{\citenamefont {Robicheaux}\ and\ \citenamefont
  {Hanson}(2002)}]{robicheaux_prl_02}%
  \BibitemOpen
  \bibfield  {author} {\bibinfo {author} {\bibfnamefont {F.}~\bibnamefont
  {Robicheaux}}\ and\ \bibinfo {author} {\bibfnamefont {J.~D.}\ \bibnamefont
  {Hanson}},\ }\href {\doibase 10.1103/PhysRevLett.88.055002} {\bibfield
  {journal} {\bibinfo  {journal} {Phys. Rev. Lett.}\ }\textbf {\bibinfo
  {volume} {88}},\ \bibinfo {pages} {055002} (\bibinfo {year}
  {2002})}\BibitemShut {NoStop}%
\bibitem [{\citenamefont {Killian}\ \emph {et~al.}(2001)\citenamefont
  {Killian}, \citenamefont {Lim}, \citenamefont {Kulin}, \citenamefont {Dumke},
  \citenamefont {Bergeson},\ and\ \citenamefont {Rolston}}]{killian_prl_01}%
  \BibitemOpen
  \bibfield  {author} {\bibinfo {author} {\bibfnamefont {T.~C.}\ \bibnamefont
  {Killian}}, \bibinfo {author} {\bibfnamefont {M.~J.}\ \bibnamefont {Lim}},
  \bibinfo {author} {\bibfnamefont {S.}~\bibnamefont {Kulin}}, \bibinfo
  {author} {\bibfnamefont {R.}~\bibnamefont {Dumke}}, \bibinfo {author}
  {\bibfnamefont {S.~D.}\ \bibnamefont {Bergeson}}, \ and\ \bibinfo {author}
  {\bibfnamefont {S.~L.}\ \bibnamefont {Rolston}},\ }\href {\doibase
  10.1103/PhysRevLett.86.3759} {\bibfield  {journal} {\bibinfo  {journal}
  {Phys. Rev. Lett.}\ }\textbf {\bibinfo {volume} {86}},\ \bibinfo {pages}
  {3759} (\bibinfo {year} {2001})}\BibitemShut {NoStop}%
\bibitem [{\citenamefont {Guo}\ \emph {et~al.}(2010)\citenamefont {Guo},
  \citenamefont {Lu},\ and\ \citenamefont {Han}}]{guo_pre_10}%
  \BibitemOpen
  \bibfield  {author} {\bibinfo {author} {\bibfnamefont {L.}~\bibnamefont
  {Guo}}, \bibinfo {author} {\bibfnamefont {R.~H.}\ \bibnamefont {Lu}}, \ and\
  \bibinfo {author} {\bibfnamefont {S.~S.}\ \bibnamefont {Han}},\ }\href
  {\doibase 10.1103/PhysRevE.81.046406} {\bibfield  {journal} {\bibinfo
  {journal} {Phys. Rev. E}\ }\textbf {\bibinfo {volume} {81}},\ \bibinfo
  {pages} {046406} (\bibinfo {year} {2010})}\BibitemShut {NoStop}%
\bibitem [{\citenamefont {Kuzmin}\ and\ \citenamefont
  {O’Neil}(2002)}]{kuzmin_pop_02}%
  \BibitemOpen
  \bibfield  {author} {\bibinfo {author} {\bibfnamefont {S.~G.}\ \bibnamefont
  {Kuzmin}}\ and\ \bibinfo {author} {\bibfnamefont {T.~M.}\ \bibnamefont
  {O’Neil}},\ }\href {\doibase http://dx.doi.org/10.1063/1.1497166}
  {\bibfield  {journal} {\bibinfo  {journal} {Physics of Plasmas}\ }\textbf
  {\bibinfo {volume} {9}},\ \bibinfo {pages} {3743} (\bibinfo {year}
  {2002})}\BibitemShut {NoStop}%
\bibitem [{\citenamefont {Kuzmin}\ and\ \citenamefont
  {O'Neil}(2002)}]{kuzmin_prl_02}%
  \BibitemOpen
  \bibfield  {author} {\bibinfo {author} {\bibfnamefont {S.~G.}\ \bibnamefont
  {Kuzmin}}\ and\ \bibinfo {author} {\bibfnamefont {T.~M.}\ \bibnamefont
  {O'Neil}},\ }\href {\doibase 10.1103/PhysRevLett.88.065003} {\bibfield
  {journal} {\bibinfo  {journal} {Phys. Rev. Lett.}\ }\textbf {\bibinfo
  {volume} {88}},\ \bibinfo {pages} {065003} (\bibinfo {year}
  {2002})}\BibitemShut {NoStop}%
\bibitem [{\citenamefont {Mazevet}\ \emph {et~al.}(2002)\citenamefont
  {Mazevet}, \citenamefont {Collins},\ and\ \citenamefont
  {Kress}}]{mazevet_02}%
  \BibitemOpen
  \bibfield  {author} {\bibinfo {author} {\bibfnamefont {S.}~\bibnamefont
  {Mazevet}}, \bibinfo {author} {\bibfnamefont {L.~A.}\ \bibnamefont
  {Collins}}, \ and\ \bibinfo {author} {\bibfnamefont {J.~D.}\ \bibnamefont
  {Kress}},\ }\href {\doibase 10.1103/PhysRevLett.88.055001} {\bibfield
  {journal} {\bibinfo  {journal} {Phys. Rev. Lett.}\ }\textbf {\bibinfo
  {volume} {88}},\ \bibinfo {pages} {055001} (\bibinfo {year}
  {2002})}\BibitemShut {NoStop}%
\bibitem [{\citenamefont {Fortov}\ and\ \citenamefont
  {Iakubov}(2000)}]{fortov_2000}%
  \BibitemOpen
  \bibfield  {author} {\bibinfo {author} {\bibfnamefont {V.~E.}\ \bibnamefont
  {Fortov}}\ and\ \bibinfo {author} {\bibfnamefont {I.}~\bibnamefont
  {Iakubov}},\ }\href {https://books.google.com/books?id=vFuZGoZsO2IC} {\emph
  {\bibinfo {title} {The Physics of Non-ideal Plasma}}}\ (\bibinfo  {publisher}
  {World Scientific},\ \bibinfo {year} {2000})\BibitemShut {NoStop}%
\bibitem [{\citenamefont {Klimontovich}(1967)}]{klimontovich1967}%
  \BibitemOpen
  \bibfield  {author} {\bibinfo {author} {\bibfnamefont {I.}~\bibnamefont
  {Klimontovich}},\ }\href {https://books.google.com/books?id=VW95AAAAIAAJ}
  {\emph {\bibinfo {title} {The statistical theory of non-equilibrium processes
  in a plasma}}},\ International series of monographs in natural philosophy\
  (\bibinfo  {publisher} {M.I.T. Press},\ \bibinfo {year} {1967})\BibitemShut
  {NoStop}%
\bibitem [{\citenamefont {Evans}\ and\ \citenamefont
  {Holian}(1985)}]{evans_85}%
  \BibitemOpen
  \bibfield  {author} {\bibinfo {author} {\bibfnamefont {D.~J.}\ \bibnamefont
  {Evans}}\ and\ \bibinfo {author} {\bibfnamefont {B.~L.}\ \bibnamefont
  {Holian}},\ }\href {\doibase http://dx.doi.org/10.1063/1.449071} {\bibfield
  {journal} {\bibinfo  {journal} {The Journal of Chemical Physics}\ }\textbf
  {\bibinfo {volume} {83}},\ \bibinfo {pages} {4069} (\bibinfo {year}
  {1985})}\BibitemShut {NoStop}%
\bibitem [{\citenamefont {Deutsch}(1977)}]{deutsch_77}%
  \BibitemOpen
  \bibfield  {author} {\bibinfo {author} {\bibfnamefont {C.}~\bibnamefont
  {Deutsch}},\ }\href {\doibase http://dx.doi.org/10.1016/0375-9601(77)90111-6}
  {\bibfield  {journal} {\bibinfo  {journal} {Physics Letters A}\ }\textbf
  {\bibinfo {volume} {60}},\ \bibinfo {pages} {317 } (\bibinfo {year}
  {1977})}\BibitemShut {NoStop}%
\bibitem [{\citenamefont {Kelbg}(1963)}]{kelbg_63}%
  \BibitemOpen
  \bibfield  {author} {\bibinfo {author} {\bibfnamefont {G.}~\bibnamefont
  {Kelbg}},\ }\href {\doibase 10.1002/andp.19634670703} {\bibfield  {journal}
  {\bibinfo  {journal} {Annalen der Physik}\ }\textbf {\bibinfo {volume}
  {467}},\ \bibinfo {pages} {354} (\bibinfo {year} {1963})}\BibitemShut
  {NoStop}%
\bibitem [{\citenamefont {Taylor}(2004)}]{taylor_04}%
  \BibitemOpen
  \bibfield  {author} {\bibinfo {author} {\bibfnamefont {J.}~\bibnamefont
  {Taylor}},\ }\href {https://books.google.com/books?id=PqHmkQEACAAJ} {\emph
  {\bibinfo {title} {Classical Mechanics}}}\ (\bibinfo  {publisher} {University
  Science Books},\ \bibinfo {year} {2004})\BibitemShut {NoStop}%
\bibitem [{\citenamefont {Fletcher}\ \emph {et~al.}(2007)\citenamefont
  {Fletcher}, \citenamefont {Zhang},\ and\ \citenamefont
  {Rolston}}]{fletcher_prl_07}%
  \BibitemOpen
  \bibfield  {author} {\bibinfo {author} {\bibfnamefont {R.~S.}\ \bibnamefont
  {Fletcher}}, \bibinfo {author} {\bibfnamefont {X.~L.}\ \bibnamefont {Zhang}},
  \ and\ \bibinfo {author} {\bibfnamefont {S.~L.}\ \bibnamefont {Rolston}},\
  }\href {\doibase 10.1103/PhysRevLett.99.145001} {\bibfield  {journal}
  {\bibinfo  {journal} {Phys. Rev. Lett.}\ }\textbf {\bibinfo {volume} {99}},\
  \bibinfo {pages} {145001} (\bibinfo {year} {2007})}\BibitemShut {NoStop}%
\bibitem [{\citenamefont {Hill}(2012)}]{hill_2012}%
  \BibitemOpen
  \bibfield  {author} {\bibinfo {author} {\bibfnamefont {T.}~\bibnamefont
  {Hill}},\ }\href {https://books.google.com/books?id=gkldAwAAQBAJ} {\emph
  {\bibinfo {title} {An Introduction to Statistical Thermodynamics}}},\ Dover
  Books on Physics\ (\bibinfo  {publisher} {Dover Publications},\ \bibinfo
  {year} {2012})\BibitemShut {NoStop}%
\bibitem [{\citenamefont {Chen}(1974)}]{chen1974}%
  \BibitemOpen
  \bibfield  {author} {\bibinfo {author} {\bibfnamefont {F.}~\bibnamefont
  {Chen}},\ }\href {https://books.google.com/books?id=u8nvAAAAMAAJ} {\emph
  {\bibinfo {title} {Introduction to plasma physics}}}\ (\bibinfo  {publisher}
  {Plenum Press},\ \bibinfo {year} {1974})\BibitemShut {NoStop}%
\bibitem [{\citenamefont {Baalrud}\ and\ \citenamefont
  {Daligault}(2014)}]{baalrud_pop_14}%
  \BibitemOpen
  \bibfield  {author} {\bibinfo {author} {\bibfnamefont {S.~D.}\ \bibnamefont
  {Baalrud}}\ and\ \bibinfo {author} {\bibfnamefont {J.}~\bibnamefont
  {Daligault}},\ }\href
  {http://scitation.aip.org/content/aip/journal/pop/21/5/10.1063/1.4875282}
  {\bibfield  {journal} {\bibinfo  {journal} {Physics of Plasmas}\ }\textbf
  {\bibinfo {volume} {21}},\ \bibinfo {eid} {055707} (\bibinfo {year}
  {2014})}\BibitemShut {NoStop}%
\bibitem [{\citenamefont {Hansen}\ and\ \citenamefont
  {McDonald}(1981)}]{hansen_pra_81}%
  \BibitemOpen
  \bibfield  {author} {\bibinfo {author} {\bibfnamefont {J.~P.}\ \bibnamefont
  {Hansen}}\ and\ \bibinfo {author} {\bibfnamefont {I.~R.}\ \bibnamefont
  {McDonald}},\ }\href {\doibase 10.1103/PhysRevA.23.2041} {\bibfield
  {journal} {\bibinfo  {journal} {Phys. Rev. A}\ }\textbf {\bibinfo {volume}
  {23}},\ \bibinfo {pages} {2041} (\bibinfo {year} {1981})}\BibitemShut
  {NoStop}%
\bibitem [{\citenamefont {Lieb}(1976)}]{lieb_76}%
  \BibitemOpen
  \bibfield  {author} {\bibinfo {author} {\bibfnamefont {E.~H.}\ \bibnamefont
  {Lieb}},\ }\href {\doibase 10.1103/RevModPhys.48.553} {\bibfield  {journal}
  {\bibinfo  {journal} {Rev. Mod. Phys.}\ }\textbf {\bibinfo {volume} {48}},\
  \bibinfo {pages} {553} (\bibinfo {year} {1976})}\BibitemShut {NoStop}%
\bibitem [{\citenamefont {Plimpton}(1995)}]{Plimpton1995}%
  \BibitemOpen
  \bibfield  {author} {\bibinfo {author} {\bibfnamefont {S.}~\bibnamefont
  {Plimpton}},\ }\href {\doibase http://dx.doi.org/10.1006/jcph.1995.1039}
  {\bibfield  {journal} {\bibinfo  {journal} {Journal of Computational
  Physics}\ }\textbf {\bibinfo {volume} {117}},\ \bibinfo {pages} {1 }
  (\bibinfo {year} {1995})}\BibitemShut {NoStop}%
\bibitem [{\citenamefont {Dimonte}\ and\ \citenamefont
  {Daligault}(2008)}]{dimonte_prl_08}%
  \BibitemOpen
  \bibfield  {author} {\bibinfo {author} {\bibfnamefont {G.}~\bibnamefont
  {Dimonte}}\ and\ \bibinfo {author} {\bibfnamefont {J.}~\bibnamefont
  {Daligault}},\ }\href {\doibase 10.1103/PhysRevLett.101.135001} {\bibfield
  {journal} {\bibinfo  {journal} {Phys. Rev. Lett.}\ }\textbf {\bibinfo
  {volume} {101}},\ \bibinfo {pages} {135001} (\bibinfo {year}
  {2008})}\BibitemShut {NoStop}%
\bibitem [{\citenamefont {Plimpton}\ \emph {et~al.}(1997)\citenamefont
  {Plimpton}, \citenamefont {Pollock},\ and\ \citenamefont
  {Stevens}}]{Plimpton97}%
  \BibitemOpen
  \bibfield  {author} {\bibinfo {author} {\bibfnamefont {S.}~\bibnamefont
  {Plimpton}}, \bibinfo {author} {\bibfnamefont {R.}~\bibnamefont {Pollock}}, \
  and\ \bibinfo {author} {\bibfnamefont {M.}~\bibnamefont {Stevens}},\ }in\
  \href@noop {} {\emph {\bibinfo {booktitle} {Proceedings of the Eighth SIAM
  Conference on Parallel Processing for Scientific Computing}}}\ (\bibinfo
  {year} {1997})\BibitemShut {NoStop}%
\bibitem [{\citenamefont {Glosli}\ \emph {et~al.}(2008)\citenamefont {Glosli},
  \citenamefont {Graziani}, \citenamefont {More}, \citenamefont {Murillo},
  \citenamefont {Streitz}, \citenamefont {Surh}, \citenamefont {Benedict},
  \citenamefont {Hau-Riege}, \citenamefont {Langdon},\ and\ \citenamefont
  {London}}]{glosli_pre_08}%
  \BibitemOpen
  \bibfield  {author} {\bibinfo {author} {\bibfnamefont {J.~N.}\ \bibnamefont
  {Glosli}}, \bibinfo {author} {\bibfnamefont {F.~R.}\ \bibnamefont
  {Graziani}}, \bibinfo {author} {\bibfnamefont {R.~M.}\ \bibnamefont {More}},
  \bibinfo {author} {\bibfnamefont {M.~S.}\ \bibnamefont {Murillo}}, \bibinfo
  {author} {\bibfnamefont {F.~H.}\ \bibnamefont {Streitz}}, \bibinfo {author}
  {\bibfnamefont {M.~P.}\ \bibnamefont {Surh}}, \bibinfo {author}
  {\bibfnamefont {L.~X.}\ \bibnamefont {Benedict}}, \bibinfo {author}
  {\bibfnamefont {S.}~\bibnamefont {Hau-Riege}}, \bibinfo {author}
  {\bibfnamefont {A.~B.}\ \bibnamefont {Langdon}}, \ and\ \bibinfo {author}
  {\bibfnamefont {R.~A.}\ \bibnamefont {London}},\ }\href {\doibase
  10.1103/PhysRevE.78.025401} {\bibfield  {journal} {\bibinfo  {journal} {Phys.
  Rev. E}\ }\textbf {\bibinfo {volume} {78}},\ \bibinfo {pages} {025401}
  (\bibinfo {year} {2008})}\BibitemShut {NoStop}%
\end{thebibliography}%
%
\end{document}